\documentclass[showpacs,aps,prd,reprint,superscriptaddress,nofootinbib,longbibliography,preprintnumbers]{revtex4-1}
\usepackage[colorlinks=true, pdfstartview=FitV, linkcolor=magenta,citecolor=blue, urlcolor=magenta,
bookmarks=true, bookmarksnumbered=true, breaklinks]{hyperref}
\usepackage[dvipdfmx]{graphicx}
\usepackage{here}
\usepackage{url}
\usepackage{amsmath,amssymb,bm,color,longtable,mathrsfs,amsfonts,slashed}
\usepackage{ulem}

\newcommand{\Slash}[1]{{\ooalign{\hfil/\hfil\crcr$#1$}}}

\begin{document}
\begin{flushright}
\end{flushright}

\title{Fate of $\Sigma_c$, $\Xi_c'$ and $\Omega_c$ baryons at high temperature with chiral restoration }

\author{Daiki~Suenaga}
\email[]{suenaga.daiki.j1@f.mail.nagoya-u.ac.jp}
\affiliation{Kobayashi-Maskawa Institute for the Origin of Particles and the Universe, Nagoya University, Nagoya, 464-8602, Japan}
\affiliation{Research Center for Nuclear Physics,
Osaka University, Ibaraki 567-0048, Japan }

\author{Makoto~Oka}
\email[]{makoto.oka@riken.jp}
\affiliation{Few-body Systems in Physics Laboratory, RIKEN Nishina Center, Wako 351-0198, Japan}
\affiliation{Advanced Science Research Center, Japan Atomic Energy Agency (JAEA), Tokai 319-1195, Japan}

\date{\today}

\begin{abstract}
Masses of the singly heavy baryons (SHBs), composed of a heavy quark and a light diquark, are studied from 
the viewpoints of heavy-quark spin symmetry (HQSS) and chiral-symmetry restoration at finite temperature. We consider the light diquarks with spin-parity $J^P=0^\pm$ and $1^\pm$. Medium corrections to the SHBs are provided through the diquarks whereas the heavy quark is simply regarded as a spectator. The chiral dynamics of the diquark are described by the Nambu-Jona-Lasinio (NJL) model having (pseudo)scalar-type and (axial)vector-type four-point interactions and the six-point ones responsible for the $U(1)$ axial anomaly. The divergences are handled by means of the three-dimensional proper-time regularization with both ultraviolet and infrared cutoffs included, in order to eliminate unphysical imaginary parts. As a result, the mass degeneracies between the parity partners of all the SHBs are predicted in accordance with the chiral restoration. In particular, the HQS-doublet SHBs exhibit clear mass degeneracies due to the absence of the direct anomaly effects. We also predict a mass degeneracy of $\Sigma_c$ and $\Omega_c$ above the pseudocritical temperature $T_{\rm pc}$ of chiral restoration, which results in a peculiar mass hierarchy for positive-parity HQS-doublet SHBs where $\Xi_c'$ becomes heavier than $\Omega_c$ Besides, it is found that the decay width of $\Sigma_c\to\Lambda_c\pi$ vanishes above $T_{\rm pc}$ reflecting a closing of the threshold. The predicted modifications of masses and decay widths of the SHBs are expected to provide future heavy-ion collision experiments and lattice simulations with useful information on chiral dynamics of the diquarks. 
 \end{abstract}


\maketitle

\section{Introduction}
\label{sec:Introduction}

A diquark, a cluster made of two quarks, is known as a useful building block of hadrons. In particular, light diquarks play a significant role in determining dynamical properties of singly heavy baryons (SHBs) constructed by two light quarks ($u,d,s$) and one heavy quark ($c,b$), since the heavy quark can be regraded as a spectator due to its heavy mass from the aspect of heavy-quark spin symmetry (HQSS)~\cite{Neubert:1993mb,manohar2000heavy}. Having focused on this advantage of diquarks, e.g., theoretical analyses based on the quark model~\cite{Copley:1979wj,Yoshida:2015tia} and the diquark--heavy-quark potential model~\cite{Santopinto:2004hw,Santopinto:2018ljf,Kim:2020imk,Kim:2021ywp,Kim:2022pyq} have been conducted.

It is well known that quantum chromodynamics (QCD) has (approximate) chiral symmetry for light flavors, and this symmetry is spontaneously broken in the vacuum. Thus, it is inevitable to take chiral dynamics into account in describing light hadrons from QCD point of view. In this regard, pions and kaons are regarded as the Nambu-Goldstone (NG) bosons associated with chiral symmetry breaking. Thanks to those chiral-symmetry aspects, coupling properties among the NG bosons and other light hadrons are delineated in a rather systematic way~\cite{cheng1994gauge,weinberg1996quantum}. In addition to the light hadrons, chiral symmetry governs the dynamics of the diquarks in SHBs too. So far, mass spectrum and decay properties of the SHBs~\cite{Ebert:1995fp,Kawakami:2018olq,Kawakami:2019hpp,Harada:2019udr,Dmitrasinovic:2020wye,Kawakami:2020sxd} including their pentaquark pictures~\cite{Suenaga:2021qri,Suenaga:2022ajn,Takada:2023evq} have been investigated based on chiral models of the diquarks.

Chiral symmetry is a continuous symmetry in terms of left-handed and right-handed quarks defined through the parity eigenvalues. Hence, we are able to examine not only the ground state but also the excited state carrying an opposite parity, i.e., the {\it parity partner} or {\it chiral partner}, in a unified way when the linear representation of chiral symmetry is adopted~\cite{cheng1994gauge,weinberg1996quantum}. The mass difference between these two states are generated by the spontaneous breakdown of chiral symmetry, and thus their masses tend to become degenerate as chiral symmetry is restored. This characteristic feature is often referred to as the {\it chiral-partner structure}~\cite{Hatsuda:1994pi}. 

Another important symmetry aspect of QCD is the nonconservation of the $U(1)$ axial charge due to topological gluon configurations, i.e., the $U(1)$ axial anomaly~\cite{tHooft:1986ooh}. For light hadrons this anomaly is essential to explain why the $\eta'$ meson is so heavy that it cannot be regarded as an NG boson~\cite{Weinberg:1975ui}. For diquarks, the direct anomaly effects were found to lead to the so-called {\it inverse mass hierarchy} of $0^-$ diquarks; the mass of $ud$ $(0^-)$ diquark can be larger than that of  $su$ $(0^-)$ [and $sd$ $(0^-)$] diquark in contrast to the naive expectation from their quark contents~\cite{Harada:2019udr}. 

One useful testing ground to explore chiral dynamics of hadrons in more detail is finite-temperature system, since in such hot medium chiral symmetry tends to be restored~\cite{Aoki:2009sc,HotQCD:2018pds}. For instance, focusing on the chiral-partner structure our understanding of the mass origin of hadrons is deepened. Moreover, this deeper understanding enables us to shed light on coupling properties among the hadrons, particularly with NG bosons, by virtue of the low-energy theorems. The mass degeneracies between the chiral partners in medium have been examined for various hadrons~\cite{Chiku:1997va,Harada:2003jx,Grahl:2011yk,GomezNicola:2013pgq,Gubler:2016djf,Suenaga:2019urn,Xu:2021lxa,Detar:1988kn,Jido:1998av,Zschiesche:2006zj,Motohiro:2015taa,Suenaga:2017wbb} including open heavy systems~\cite{Suenaga:2014sga,Sasaki:2014asa,Harada:2016uca,Suenaga:2017deu,Azizi:2019cmj}.
 
In our previous paper~\cite{Suenaga:2023tcy}, modifications of spin-$0$ diquark masses at finite temperature were investigated based on a chiral model, the Nambu-Jona-Lasinio (NJL) model, particularly focusing on the $U(1)$ axial anomaly effects and chiral restoration. As a result, the inverse mass hierarchy driven by the $U(1)$ axial anomaly was found to persist at any temperatures. Furthermore, we found that the anomaly effects delay the mass degeneracies between the chiral partners of $ud$ diquarks at high temperature.

In this paper, we extend the previous work by taking spin-$1$ diquarks into account and present comprehensively predictions of the mass corrections of the diquarks in medium. In terms of multiplets of HQSS, the spin-$0$ diquarks and spin-$1$ diquarks lead to HQS-singlet SHBs with $J=\frac{1}{2}$ and HQS-doublet ones with $J=\frac{1}{2}$ and $\frac{3}{2}$, respectively~\cite{Falk:1991nq}. Therefore, we exhibit the mass modifications in terms of these SHBs as appropriate observables in order to provide future heavy-ion collision experiments and lattice simulations with useful information on chiral properties of the diquarks. 

In Refs.~\cite{Kim:2021ywp,Kim:2022pyq}, it was predicted that the mass of $\Sigma_c$ baryon becomes smaller than that of $\Lambda_c$ baryon within the diquark--heavy-quark potential description as chiral symmetry is restored. This peculiar mass ordering were, however, obtained by simply reducing the value of chiral condensates, the order parameter of chiral-symmetry breaking, although in more realistic situation medium effects give rise to other corrections upon the SHB masses. Thus, in the present work we also focus on the mass hierarchy of $\Sigma_c$ and $\Lambda_c$ and the fate of $\Sigma_c\to\Lambda_c\pi$ decays at high temperature paying particular attentions to the threshold value.

This paper is organized as follows. In Sec.~\ref{sec:Model} we introduce our NJL model toward describing the dynamics of spin-$0$ and spin-$1$ diquarks. In Sec.~\ref{sec:BSEquation} and Sec.~\ref{sec:Regularization}, we explain our method to evaluate the diquark masses and to regularize the divergences, respectively. Based on these procedures, numerical results on the SHB masses at finite temperature are presented in Sec.~\ref{sec:Numerical}. Some discussions on the mass hierarchy of the spin-$1$ diquarks are provided in Sec.~\ref{sec:DiscussionDiquark}. Moreover, the fate of the decay width of $\Sigma_c\to\Lambda_c\pi$ at high temperature is examined in Sec.~\ref{sec:DecaySigmaC}. Then, finally we conclude the present study in Sec.~\ref{sec:Conclusion}.

\section{Model construction}
\label{sec:Model}

\subsection{NJL Lagrangian}
\label{sec:NJLLagrangian}

This paper is devoted to examining the mass corrections of low-lying HQS-singlet and HQS-doublet SHBs at high temperature based on a chiral model of the composing diquarks. To this aim, in this subsection we introduce our NJL model to describe both the spin-$0$ and spin-$1$ diquarks.

Our NJL model takes the following form:
\begin{eqnarray} 
{\cal L}_{\rm NJL} = {\cal L}_{2q} + {\cal L}^{S}_{4q} + {\cal L}^{V}_{4q} + {\cal L}_{6q}^{\rm anom.} . \label{LNJL}
\end{eqnarray}
The first term ${\cal L}_{2q}$ includes the kinetic and mass terms of constituent quarks
\begin{eqnarray}
{\cal L}_{2q} = \bar{\psi}(i\Slash{\partial}-{\cal M})\psi ,
\end{eqnarray}
where $\psi=(u,d,s)$ is the quark triplet and ${\cal M}$ is the $3\times3$ quark mass matrix. In the present work we assume the $SU(2)$ isospin symmetry so that ${\cal M}={\rm diag}(m_q,m_q,m_s)$.

The second and third pieces ${\cal L}_{4q}^S$ and ${\cal L}_{4q}^V$ in Eq.~(\ref{LNJL}) represent the (pseudo)scalar-type and (axial)vector-type four-point interactions, respectively, which are given by~\cite{Buballa:2003qv}
\begin{eqnarray}
&& {\cal L}^S_{4q} = G_S\big[(\bar{\psi}\lambda_f^X\psi)^2 + (\bar{\psi}i\gamma_5\lambda_f^X\psi)^2\big] \nonumber\\
&& + H_S\Big[|\psi^TC\lambda_f^A\lambda_c^{A'}\psi|^2 + |\psi^TC\gamma_5\lambda_f^A\lambda_c^{A'}\psi|^2 \Big]  , \label{LFourS}
\end{eqnarray}
and 
\begin{eqnarray}
{\cal L}^V_{4q} = - H_V\left[|\psi^TC\gamma^\mu\lambda_f^S\lambda_c^{A'}\psi|^2 + |\psi^TC\gamma^\mu\gamma_5\lambda_f^{A}\lambda_c^{A'}\psi |^2 \right] .\nonumber\\
\label{LFourV}
\end{eqnarray}
In these equations, $G_S$ is a coupling constant controlling the formation of spin-$0$ mesons, while $H_S$ and $H_V$ are those for spin-$0$ diquarks and spin-$1$ diquarks. $C=i\gamma^2\gamma^0$ is the charge-conjugation matrix. Besides, $\lambda^X_f$ and $\lambda^X_c$ are Hermitian $3\times3$ matrices in flavor and color spaces. Then, the superscript $X$ runs over $X=0$ - $8$ with $\lambda_{f(c)}^{X=0} = (1/\sqrt{3}){\bm 1}$ and $\lambda_{f(c)}^{X=1-8}$ being the Gell-Mann matrices. The indices $A$ ($A'$) and $S$ ($S'$) take only the antisymmetric and symmetric parts of $X$, respectively, more explicitly, $A\, (A')=2,5,7$ and $S\, (S')=0,1,3,4,6,8$. We note that these selections for diquark channels are derived from the Pauli principle of the quark bilinears, with which the allowed operators and quantum numbers are summarized in Table~\ref{tab:Bilinear}. We also note that the spin-$1$ meson channels which could correct the temperature dependences of the pion decay constant are abbreviated in our present model so as to focus on the mass spectrum of the diquarks, since such effects are expected to be comparably suppressed. On the other hand, the spin-$0$ meson coupling is necessary to determine the chiral properties in both the vacuum and high temperature.

 On the other hand, we have assumed that 

\begin{table}[htbp]
\begin{center}
  \begin{tabular}{c|cccc} \hline
Operator & $J^P$ & Flavor & Color & $^{2S+1}L_J$\\   \hline 
$\psi^TC\gamma_5\lambda_f^A\lambda_c^{A'}\psi$ & $0^+$ & $\bar{\bm 3}_f$ & $\bar{\bm 3}_c$ & $^1S_0$ \\
$\psi^TC\lambda_f^A\lambda_c^{A'}\psi$ & $0^-$ & $\bar{\bm 3}_f$ & $\bar{\bm 3}_c$ & $^3P_0$ \\
$\psi^TC\gamma^\mu\lambda_f^S\lambda_c^{A'}\psi$ & $1^+$ & ${\bm 6}_f$ & $\bar{\bm 3}_c$ & $^3S_1$ \\
$\psi^TC\gamma^\mu\gamma_5\lambda_f^A\lambda_c^{A'}\psi$ & $1^-$ & $\bar{\bm 3}_f$ & $\bar{\bm 3}_c$ & $^3P_1$ \\ \hline
 \end{tabular}
\caption{Each quantum number of the spin-$0$ and spin-$1$ diquark operators incorporated in the present NJL model.}
\label{tab:Bilinear}
\end{center}
\end{table}

The interactions~(\ref{LFourS}) and~(\ref{LFourV}) are constructed in a chiral symmetric way. To see this, we introduce the following quark composite operators whose flavor indices are uncontracted:
\begin{eqnarray}
\phi_{ij} &=& (\bar{\psi}_{R})^a_j(\psi_{L})^a_i  ,\nonumber\\
(\eta_L)_i^a &=& \epsilon_{ijk}\epsilon^{abc}(\psi_L^T)_j^bC(\psi_L)_k^c , \nonumber\\
(\eta_R)_i^a &=& \epsilon_{ijk}\epsilon^{abc}(\psi_R^T)_j^bC(\psi_R)_k^c , \nonumber\\
\tilde{\eta}_{ij}^{a,\mu} &=& \epsilon^{abc}\psi_{R,i}^{T,b}C\gamma^\mu \psi_{L,j}^c , \label{OperatorsChiral}
\label{Bilinears}
\end{eqnarray}
with $\psi_{R(L)} = \frac{1\pm\gamma_5}{2}\psi$ being the right-handed (left-handed) quark field, while $i,j,\cdots$ and $a,b,\cdots$ denote the flavor and color indices, respectively. Since the quark field transforms as $\psi_{R(L)} \to g_{R(L)}\psi_{R(L)}$ with $g_{R(L)}\in U(3)_{R(L)}$ under $U(3)_R\times U(3)_L$ chiral transformation, the chiral transformation laws of the bilinears in Eq.~(\ref{Bilinears}) read
\begin{eqnarray}
&& \phi \to g_L\phi g_R^\dagger \ , \ \   \tilde{\eta}^{a,\mu} \to g_R\tilde{\eta}^{a,\mu}g_L^T  , \nonumber\\
&&   \eta_L^a \to g^*_L\eta_L^a\ , \ \  \eta_R^a \to g^*_R\eta_R^a . \label{FieldTrans}
\end{eqnarray}
Here, using the identities
\begin{eqnarray}
&& {\rm tr}[\phi^\dagger\phi] = \frac{1}{8}\Big[(\bar{\psi}\lambda_f^X\psi)^2+(\bar{\psi}i\gamma_5\lambda_f^X\psi)^2\Big]  , \nonumber\\
&&\eta_L^\dagger\eta_L + \eta_R^\dagger\eta_R = \frac{1}{2}\Big[|\psi^TC\lambda_f^A\lambda_c^{A'}\psi|^2  + |\psi^TC\gamma_5\lambda_f^A\lambda_c^{A'}\psi|^2 \Big] , \nonumber\\
&&{\rm tr}[\tilde{\eta}^\dagger_\mu\tilde{\eta}^\mu]  = \frac{1}{8}\Big[|\psi^TC\gamma^\mu\lambda_f^S\lambda_c^{A'}\psi|^2  +| \psi^TC\gamma_5\gamma^\mu\lambda_f^{A}\lambda_c^{A'}\psi|^2\Big] ,\nonumber\\
\end{eqnarray}
one can rewrite the four-point interactions~(\ref{LFourS}) and~(\ref{LFourV}) into
\begin{eqnarray}
{\cal L}_{4q}^S =8G_S{\rm tr}[\phi^\dagger\phi] +2H_S(\eta_L^\dagger\eta_L+\eta_R^\dagger\eta_R) , \label{L4qSChiral}
\end{eqnarray}
and
\begin{eqnarray}
{\cal L}_{4q}^V = 8 {\rm tr}[\tilde{\eta}^\dagger_\mu\tilde{\eta}^\mu] , \label{L4qVChiral}
\end{eqnarray}
which are indeed chiral symmetric from Eq.~(\ref{FieldTrans}).

The last contribution in Eq.~(\ref{LNJL}), $ {\cal L}_{6q}^{\rm anom.}$, is responsible for the $U(1)$ axial anomaly. That is, this term is invariant under $SU(3)_L\times SU(3)_R$ chiral transformation but violates $U(1)$ axial symmetry. In this work we employ the following interactions:
\begin{eqnarray}
{\cal L}_{6q}^{\rm anom.} &=& -8K({\rm det}\phi + {\rm det}\phi^\dagger)  \nonumber\\
&+& K'(\eta^T_L\phi\eta_R^*+\eta^T_R \phi^\dagger\eta_L^*) . \label{LSix}
\end{eqnarray}
The first term is the so-called Kobayashi-Maskawa-'tHooft (KMT) determinant interaction~\cite{Kobayashi:1970ji,Kobayashi:1971qz,tHooft:1976snw,tHooft:1976rip} while the second one describes the anomalous coupling among the mesons and diquarks~\cite{Yamamoto:2007ah,Abuki:2010jq}.

The spontaneous breaking of chiral symmetry can be expressed by generation of the vacuum expectation values (VEVs) of $\bar{\psi}\psi$. (We denote the VEVs by $\langle{\cal O}\rangle$.)
Under the $SU(2)$ isospin symmetry, these VEVs are given by $\langle\phi\rangle = {\rm diag}(\langle\bar{q}q\rangle,\langle\bar{q}q\rangle,\langle\bar{s}s\rangle)$. Then, by employing the following approximation in the four-point and six-point interactions in Eq.~(\ref{LNJL}):
\begin{eqnarray}
XY &\approx& XY+ \langle X\rangle Y+\langle Y\rangle X-\langle X\rangle\langle Y\rangle \ , \nonumber\\
XYZ &\approx& \langle X\rangle YZ + \langle Y\rangle XZ + \langle Z\rangle XY + \langle X\rangle\langle Y\rangle Z \nonumber\\
&& +\langle Y\rangle\langle Z\rangle X + \langle Z\rangle\langle X\rangle Y-2\langle X\rangle\langle Y\rangle\langle Z\rangle, \label{Approximation}
\end{eqnarray}
at the mean-field level the masses of $q$ $(=u,d)$ and $s$ quarks read
 \begin{eqnarray}
M_q &=& m_q-4G_S\langle\bar{q}q\rangle + 2K\langle\bar{q}q\rangle\langle\bar{s}s\rangle, \nonumber\\
M_s &=& m_s -4G_S\langle\bar{s}s\rangle + 2K\langle\bar{q}q\rangle^2, \label{ConsMass}
\end{eqnarray}
respectively. From this formula one can see that the anomalous $K$ interactions provide $\langle\bar{s}s\rangle$ contributions to the $u$ ($d$) quark masses while they only generate $\langle \bar{q}q\rangle$ ones to the $s$ quark mass. 

\subsection{Kernel}
\label{sec:Kernel}

The chiral symmetric interactions which is capable of generating spin-$0$ mesons and spin-$0$ and spin-$1$ diquarks have been introduced in Sec.~\ref{sec:NJLLagrangian}. Their masses are read off by pole positions of the amplitude from the corresponding composite operators~\cite{Hellstern:1997nv}. The procedure for the meson channels is already well-known, however, for the diquark channels it would not be straightforward, particularly to find appropriate kernels in computing the amplitude, due to constraints from the Pauli principle. For this reason, before moving on to the evaluation of the scattering amplitudes, in this subsection, we present proper diquark operators and rewrite the interaction Lagrangians into a useful form to find the corresponding kernels.

As for the spin-$0$ diquarks, the mass-eigenstate operators are simply given by
\begin{eqnarray}
(\eta_+)_i^a &=& \frac{1}{\sqrt{2}}(\eta_R-\eta_L)_i^a = \frac{1}{\sqrt{2}}\epsilon_{ijk}\epsilon^{abc}\psi_j^{T,b} C\gamma_5 \psi_k^c , \nonumber\\
(\eta_-)_i^a &=& \frac{1}{\sqrt{2}}(\eta_R+\eta_L)_i^a = \frac{1}{\sqrt{2}}\epsilon_{ijk}\epsilon^{abc}\psi_j^{T,b} C \psi_k^c, \label{Eta+-}
\end{eqnarray}
thus, we define
\begin{eqnarray}
\eta_{[ud]_0^+}^a &=& \frac{\epsilon^{abc}}{\sqrt{2}}\left(u^{T,b}C\gamma_5d^c-d^{T,b}C\gamma_5 u^c\right) , \nonumber\\
\eta_{[su]_0^+}^a &=& \frac{\epsilon^{abc}}{\sqrt{2}}\left(s^{T,b}C\gamma_5u^c-u^{T,b}C\gamma_5 s^c\right) , \nonumber\\
\eta_{[sd]_0^+}^a &=& \frac{\epsilon^{abc}}{\sqrt{2}}\left(s^{T,b}C\gamma_5d^c-d^{T,b}C\gamma_5 s^c\right) , 
\end{eqnarray}
and
\begin{eqnarray}
\eta_{[ud]_0^-}^a &=& \frac{\epsilon^{abc}}{\sqrt{2}}\left(u^{T,b}Cd^c-d^{T,b}C u^c\right) , \nonumber\\
\eta_{[su]_0^-}^a &=& \frac{\epsilon^{abc}}{\sqrt{2}}\left(s^{T,b}Cu^c-u^{T,b}C s^c\right) , \nonumber\\
\eta_{[sd]_0^-}^a &=& \frac{\epsilon^{abc}}{\sqrt{2}}\left(s^{T,b}Cd^c-d^{T,b}C s^c\right) , 
\end{eqnarray}
to refer to the scalar and pseudoscalar diquarks. Using these operators the interaction terms in Eqs.~(\ref{L4qSChiral}) and~(\ref{LSix}) which are relevant to the scattering amplitude within the approximation~(\ref{Approximation}) read
\begin{eqnarray}
{\cal L}_{\rm int}^{S} &=&  {\cal K}_{[qq]_0^+} |\eta^a_{[ud]_0^+}|^2 +{\cal K}_{[sq]_0^+} \left( |\eta^a_{[su]_0^+}|^2 + |\eta^a_{[sd]_0^+}|^2 \right) \nonumber\\
&+& {\cal K}_{[qq]_0^-} |\eta^a_{[ud]_0^+}|^2 +{\cal K}_{[sq]_0^-} \left( |\eta^a_{[su]_0^+}|^2 + |\eta^a_{[sd]_0^+}|^2 \right) , \nonumber\\
\end{eqnarray}
where the each kernel is defined by
\begin{eqnarray}
{\cal K}_{[qq]_0^\pm} &\equiv& 2H_S\mp\frac{K'}{2}\langle\bar{s}s\rangle , \nonumber\\
{\cal K}_{[sq]_0^\pm} &\equiv& 2H_S\mp\frac{K'}{2}\langle\bar{q}q\rangle . \label{Kernel0}
\end{eqnarray}
The sign of $K'$ term incorporating the chiral condensates differs for positive and negative diquarks which reflects the chiral-partner structure; the kernels converge to the identical value when chiral symmetry is completely restored. Besides, the $\langle\bar{s}s\rangle$ and $\langle\bar{q}q\rangle$ contribute to the kernels for $ud$ and $su$ ($sd$) diquarks, respectively, hence, again one can understand that the anomalous contributions generate the ``flavor-mixing'' structure with respect to the kernels. These properties significantly affect the attractive force to form the diquarks and their masses.

The spin-$1$ diquarks are rather complicated since the flavor structure is correlated with the parity. For this reason, we separate the operator $\tilde{\eta}_{ij}^{a,\mu}$ in Eq.~(\ref{OperatorsChiral}) as
\begin{eqnarray}
\tilde{\eta}_{ij}^{a,\mu} = \frac{1}{2}(\tilde{\eta}_{\bm 6})_{ij}^{a,\mu} + \frac{1}{2}(\tilde{\eta}_{\bar{\bm 3}})_{ij}^{a,\mu} ,  \label{EtaTildeDec}  
\end{eqnarray}
where the flavor sextet and antitriplet matrices are given by
\begin{eqnarray}
&&\tilde{\eta}_{\bm 6}^{a,\mu}= \left(
\begin{array}{ccc}
\eta_{\{uu\}_1^+} & \frac{1}{\sqrt{2}}\eta_{\{ud\}_1^+} & \frac{1}{\sqrt{2}}\eta_{\{su\}_1^+} \\
\frac{1}{\sqrt{2}}\eta_{\{ud\}_1^+} & \eta_{\{dd\}_1^+} & \frac{1}{\sqrt{2}}\eta_{\{sd\}_1^+} \\
\frac{1}{\sqrt{2}}\eta_{\{su\}_1^+} & \frac{1}{\sqrt{2}}\eta_{\{sd\}_1^+} & \eta_{\{ss\}_1^+} \\
\end{array}
\right)^{a,\mu} , \nonumber\\
&& \tilde{\eta}_{\bar{\bm 3}}^{a,\mu} = \left(
\begin{array}{ccc}
0 & \frac{1}{\sqrt{2}}\eta_{[ud]_1^-} & \frac{1}{\sqrt{2}}\eta_{[su]_1^-} \\
-\frac{1}{\sqrt{2}}\eta_{[ud]_1^-} &0& \frac{1}{\sqrt{2}}\eta_{[sd]_1^+} \\
-\frac{1}{\sqrt{2}}\eta_{[su]_1^-}&- \frac{1}{\sqrt{2}}\eta_{[sd]_1^+} & 0 \\
\end{array}
\right)^{a,\mu} , \nonumber\\
\end{eqnarray}
with the axialvector- and vector-diquark operators being
\begin{eqnarray}
(\tilde{\eta}_{\{uu\}_1^+})^{a,\mu} &=& \epsilon^{abc}u^{T,b}C\gamma^\mu u^c , \nonumber\\
(\tilde{\eta}_{\{dd\}_1^+})^{a,\mu} &=& \epsilon^{abc}d^{T,b}C\gamma^\mu d^c,  \nonumber\\
(\tilde{\eta}_{\{ss\}_1^+})^{a,\mu} &=& \epsilon^{abc}s^{T,b}C\gamma^\mu s^c,  \nonumber\\
(\tilde{\eta}_{\{ud\}_1^+})^{a,\mu} &=& \frac{\epsilon^{abc}}{\sqrt{2}}\left(u^{T,b}C\gamma^\mu d^c + d^{T,b}C\gamma^\mu u^c \right) ,  \nonumber\\
(\tilde{\eta}_{\{su\}_1^+})^{a,\mu} &=& \frac{\epsilon^{abc}}{\sqrt{2}}\left(u^{T,b}C\gamma^\mu s^c + s^{T,b}C\gamma^\mu u^c \right) ,\nonumber\\
(\tilde{\eta}_{\{sd\}_1^+})^{a,\mu} &=& \frac{\epsilon^{abc}}{\sqrt{2}}\left(d^{T,b}C\gamma^\mu s^c + s^{T,b}C\gamma^\mu d^c \right) ,
\label{AVDQInt}
\end{eqnarray}
and
\begin{eqnarray}
(\tilde{\eta}_{[ud]_1^-})^{a,\mu} &=& \frac{\epsilon^{abc}}{\sqrt{2}}\left(u^{T,b}C\gamma_5 \gamma^\mu d^c - d^{T,b}C\gamma_5\gamma^\mu u^c \right) , \nonumber\\
(\tilde{\eta}_{[su]_1^-})^{a,\mu} &=& \frac{\epsilon^{abc}}{\sqrt{2}}\left(s^{T,b}C\gamma_5\gamma^\mu u^c - u^{T,b}C\gamma_5 \gamma^\mu s^c \right),  \nonumber\\
(\tilde{\eta}_{[sd]_1^-})^{a,\mu} &=& \frac{\epsilon^{abc}}{\sqrt{2}}\left(d^{T,b}C\gamma_5 \gamma^\mu s^c - s^{T,b}C\gamma_5\gamma^\mu d^c \right) ,\nonumber\\
\label{VDQInt}
\end{eqnarray}
respectively. Hence, with these operators the interaction terms are read off from Eqs.~(\ref{L4qSChiral}) and~(\ref{LSix}) as
\begin{eqnarray}
{\cal L}_{\rm int}^V &=& - {\cal K}_{1^\pm} \Big(|\tilde{\eta}^{a,\mu}_{\{uu\}_1^+}|^2 + |\tilde{\eta}^{a,\mu}_{\{dd\}_1^+}|^2 + |\tilde{\eta}^{a,\mu}_{\{ss\}_1^+}|^2 + |\tilde{\eta}^{a,\mu}_{\{su\}_1^+}|^2 \nonumber\\
&+& |\tilde{\eta}^{a,\mu}_{\{sd\}_1^+}|^2 + |\tilde{\eta}^{a,\mu}_{[ud]_1^-}|^2  + |\tilde{\eta}^{a,\mu}_{[su]_1^-}|^2  + |\tilde{\eta}^{a,\mu}_{[sd]_1^-}|^2 \Big) , \label{LIntV}
\end{eqnarray}
where the corresponding kernel
\begin{eqnarray}
{\cal K}_{1^\pm} \equiv 2H_V \label{Kernel1}
\end{eqnarray}
is commonly applied to all the spin-$1$ channels. This universality is due to the absence of the direct anomaly effects on the spin-$1$ diquarks.

\section{Bethe-Salpeter equation}
\label{sec:BSEquation}

The proper interactions in terms of the diquark composite operators have been obtained in Sec.~\ref{sec:Kernel}. Based on them, here, we present the formula to evaluate the diquark masses. In the following analysis, the color indices will be omitted since $SU(3)$ color symmetry is trivial.

The amplitude ${\cal T}$ within the one-loop approximation follows the Bethe-Salpeter equation
\begin{eqnarray}
{\cal T} = {\cal K} + {\cal K}{\cal J}{\cal T}\ \ \leftrightarrow\ \ {\cal T} = ({\cal K}^{-1}-{\cal J})^{-1} , \label{BSEquation}
\end{eqnarray}
where ${\cal K}$ is the kernel derived in Eq.~(\ref{Kernel0}) or Eq.~(\ref{Kernel1}) for the respective channels, and ${\cal J}$ is the corresponding one-loop function. For instance, ${\cal J}$ for $[qq]_0^+$ diquark reads
\begin{eqnarray}
{\cal J}_{[qq]_0^+} = -4T\sum_m\int\frac{d^3p}{(2\pi)^3}{\rm tr}\Big[\gamma_5S_{(q)}(p')\gamma_5{S}^{\cal C}_{(q)}(p)\Big] , \label{Jud0+} 
\end{eqnarray}
within the imaginary-time formalism at arbitrary temperature $T$. In this equation, $p'=p+q$ with $p_0=i\omega_m$ ($\omega_m=(2m+1)\pi T$) and $q_0=i\bar{\omega}_n$ ($\bar{\omega}_n=2n\pi T$) being the fermionic and bosonic Matsubara frequencies, respectively~\cite{kapusta2006finite}. The quark propagator $S_{(q)}(p)$ is defined by the Fourier transformation of $\langle 0|{\rm T}u(x)\bar{u}(0)|0\rangle$ (or $\langle 0|{\rm T}d(x)\bar{d}(0)|0\rangle$) which is given by
\begin{eqnarray}
S_{(q)}(p) = \frac{i}{\Slash{p}-M_q} = i\sum_{\zeta={\rm p},{\rm a}}\frac{\Lambda^{(q)}_\zeta({\bm p})}{p_0-E^{(q)}_{\bm p}}\gamma_0, \label{Propagator}
\end{eqnarray}
with $E_{\bm p}^{(q)} = \sqrt{{\bm p}^2+M_q^2}$, where
\begin{eqnarray}
\Lambda^{(q)}_{\rm p}({\bm p}) &=& \frac{E_{\bm p}^{(q)}+(M_q\gamma_0+{\bm p}\cdot{\bm \alpha})}{2E^{(q)}_{\bm p}}, \nonumber\\
\Lambda^{(q)}_{\rm a}({\bm p}) &=& \frac{E_{\bm p}^{(q)}-(M_q\gamma_0+{\bm p}\cdot{\bm \alpha})}{2E^{(q)}_{\bm p}},
\end{eqnarray}
(${\bm \alpha}=\gamma_0{\bm \gamma}$) are the projection operators for positive-energy and negative-energy components. Meanwhile, $S_{(q)}^{\cal C}(p)$ in Eq.~(\ref{Jud0+}) is the charge-conjugated quark propagator defined by the Fourier transform of $\langle0|{\rm T}u^{\cal C}(x)\bar{u}^{\cal C}(0)|0\rangle$ (or $\langle0|{\rm T}d^{\cal C}(x)\bar{d}^{\cal C}(0)|0\rangle$) with $u^{\cal C}=C\bar{u}^T$ ($d^{\cal C}=C\bar{d}^T$) being the charge-conjugated field. Using this definition one can easily find $S^{\cal C}_{(q)}(p)$ coincides with $S_{(q)}(p)$:
\begin{eqnarray}
S^{\cal C}_{(q)}(p) = S_{(q)}(p) . \label{SaNdSC} 
 \end{eqnarray}
This coincidence stems from the charge-conjugation symmetry of the system as long as we do not take chemical potentials into account. 

After taking the Dirac trace in Eq.~(\ref{Jud0+}) with Eq.~(\ref{Propagator}), the loop function turns into
\begin{eqnarray}
{\cal J}_{[qq]_0^+} &=& -4\int\frac{d^3p}{(2\pi)^3}  \sum_{\zeta',\zeta={\rm p},{\rm a}} T^{\zeta'\zeta}_{[qq]_0^+} \nonumber\\
&\times& T\sum_m\frac{1}{(p_0'-\eta_{\zeta'}E^{(q)}_{{\bm p}'})(p_0-\eta_\zeta E^{(q)}_{\bm p})}  , \label{JudRed}
\end{eqnarray}
with $\eta_{{\rm p}/{\rm a}}=\pm1$, where the kinetic part
\begin{eqnarray}
T^{\zeta'\zeta}_{[qq]_0^+} &=& {\rm tr}[\gamma_5\Lambda_{\zeta'}^{(q)}({\bm p}')\gamma_5\Lambda_{\zeta}^{(q)}({\bm p})] \nonumber\\
&=& \frac{E_{{\bm p}'}^{(q)}E_{\bm p}^{(q)}+\eta_{\zeta'}\eta_\zeta({\bm p}'\cdot{\bm p}+M_q^2)}{E_{{\bm p}'}^{(q)}E_{\bm p}^{(q)}}
\end{eqnarray}
represents the spin couplings among the quarks and diquark. The information of thermal excitations is described by the second line of Eq.~(\ref{JudRed}) with the Matsubara summation. The summation in Eq.~(\ref{JudRed}) is straightforwardly performed with the help of the identity~\cite{kapusta2006finite}
\begin{eqnarray}
T\sum_m\frac{1}{(i\omega_m+i\bar{\omega}_n-\epsilon')(i\omega_m-\epsilon)} = \frac{f_F(\epsilon)-f_F(\epsilon')}{i\bar{\omega}_n-\epsilon'+\epsilon}\ ,  
\label{MatsubaraSum}
\end{eqnarray}
to express the thermal contributions in terms of the Fermi-Dirac distribution function
\begin{eqnarray}
f_F(\epsilon) = \frac{1}{{\rm e}^{\epsilon/T}+1} .
\end{eqnarray}
Then, after the analytic continuation to the real time, finally one can express ${\cal J}_{[qq]_0^+} $ in the following useful way:
\begin{eqnarray}
{\cal J}_{[qq]_0^+} &=& -4\int\frac{d^3p}{(2\pi)^3}\sum_{\zeta',\zeta={\rm p}, {\rm a}}\left(1+\frac{\eta_{\zeta'}\eta_\zeta({\bm p}'\cdot{\bm p}+M_q^2)}{E_{{\bm p}'}^{(q)}E_{\bm p}^{(q)}}\right) \nonumber\\
&\times&\frac{1-f_F\left(\eta_{\zeta}E_{\bm p}^{(q)}\right)-f_F\left(\eta_{\zeta'}E_{{\bm p}'}^{(q)}\right)}{q_0-\eta_{\zeta'}E_{{\bm p}'}^{(q)}-\eta_\zeta E_{\bm p}^{(q)}} . \label{Jud0+Comp}
\end{eqnarray}
From this loop function together with the kernel~(\ref{Kernel0}), the scattering amplitude leading to the $[qq]_0^+$ diquark is evaluated from the Bethe-Salpeter equation~(\ref{BSEquation}). Then, finally the mass of $[qq]_0^+$ diquark is numerically obtained by finding the pole position of the amplitude at rest frame ${\bm q}={\bm 0}$. It should be noted that the ultraviolet (UV) divergences in Eq.~(\ref{Jud0+Comp}) are regularized by inserting a regularization function as will be explained in Sec.~\ref{sec:Regularization}.

In a similar way, loop functions of the spin-$1$ diquarks, e.g., $\{qq\}_1^+$ diquark: ${\cal J}_{\{qq\}_1^+}$, can be evaluated. That is, from the interaction Lagrangian~(\ref{LIntV}) and the appropriate operator~(\ref{AVDQInt}), the loop function ${\cal J}_{\{qq\}_1^+}$ reads
\begin{eqnarray}
{\cal J}^{\mu\nu}_{\{qq\}_{1}^+} &=&  4T\sum_m\int\frac{d^3p}{(2\pi)^3} {\rm tr}\Big[ \gamma^\mu S_{(q)}(p')\gamma^\nu S_{(q)}^{\cal C}(p) \Big]  . \nonumber\\ \label{Jud1+}
\end{eqnarray}
Here, Using the charge-conjugation symmetry for the propagator in Eq.~(\ref{SaNdSC}) and calculating the trace of the Dirac matrices, one can get
\begin{eqnarray}
{\cal J}^{\mu\nu}_{\{qq\}_{1}^+} &=&  -4\int\frac{d^3p}{(2\pi)^3}\sum_{\zeta',\zeta={\rm p,a}}T^{\zeta'\zeta,\mu\nu}_{\{qq\}_1^+} \nonumber\\
&\times& \frac{1-f_F\left(\eta_{\zeta}E_{\bm p}^{(q)}\right)-f_F\left(\eta_{\zeta'}E_{{\bm p}'}^{(q)}\right)}{q_0-\eta_{\zeta'}E_{{\bm p}'}^{(q)}-\eta_\zeta E_{\bm p}^{(q)}} 
\end{eqnarray}
with the summation formula~(\ref{MatsubaraSum}), where the kinetic part is of the form
\begin{eqnarray}
T^{\zeta'\zeta,\mu\nu}_{\{qq\}_1^+} &=&  \frac{1}{E_{{\bm p}'}^{(q)}E_{\bm p}^{(q)}} \Big[E_{{\bm p}'}^{(q)}E_{\bm p}^{(q)}(2g^{\mu0}g^{\nu0}-g^{\mu\nu}) \nonumber\\
&+& \eta_\zeta E^{(q)}_{{\bm p}'}p^j(g^{\mu0}g^{\nu j}+g^{\mu j}g^{\nu0}) - \eta_{\zeta'}E^{(q)}_{\bm p}p'^i \nonumber\\
&\times&(g^{\mu i}g^{\nu0} + g^{\mu0}g^{\nu i})-\eta_{\zeta'}\eta_\zeta p'^ip^j(g^{\mu i}g^{\nu j}\nonumber\\
&-& g^{\mu\nu}g^{ij}+g^{\mu j}g^{i\nu})-\eta_{\zeta'}\eta_\zeta M_q^2 g^{\mu\nu}\Big] .
\end{eqnarray}
From these equations one sees that ${\cal J}^{00}_{\{qq\}_{1}^+}$, ${\cal J}^{0j}_{\{qq\}_{1}^+}$ and ${\cal J}^{ij}_{\{qq\}_{1}^+}$ differ from each other, reflecting the violation of Lorentz covariance, even at zero temperature due to the present three-dimensional integral form.

In the present study, we only focus on the mass of diquarks defined by pole positions of the scattering amplitudes at rest: ${\bm q}={\bm 0}$. In this limit, only the spatial components of the spin-$1$ diquarks become physical so that only the ${\cal J}_{\{qq\}_1^+}^{ij}$ contributes to mass evaluation. Moreover, when ${\bm q}={\bm 0}$, ${\cal J}_{\{qq\}_1^+}^{ij}$ must be proportional to $\delta^{ij}$, hence, the matrix structure of the Bethe-Salpeter equation~(\ref{BSEquation}) is reduced to a handleable form.

In a similar manner, all the one-loop functions are evaluated within three-dimensional integrals that include UV divergences. They are summarized in Appendix~\ref{sec:LoopFunctions}. It is important to see that the scalar diquark and the vector diquark may mix through derivative couplings when the flavor structure is identical. Although we take ${\bm q}\to{\bm 0}$ limit, such mixing effects survive since the zeroth component $q^0$ of the momentum is not vanishing. In other words, for instance, ${\cal J}_{[qq]_0^+}$ and ${\cal J}_{[qq]_1^-}^{00}$ are correlated by the mixed loop function ${\cal J}_{\{qq\}_0^+[ud]_1^-}^{0}\propto q^0$. Therefore, although the ${\cal J}_{[qq]_1^-}^{00}$ decouples from evaluating the $[qq]_1^-$ diquark mass with ${\bm q}\to{\bm 0}$, it provides corrections to the $[qq]_0^+$ mass. Such processes are also explicitly shown in Appendix~\ref{sec:LoopFunctions}

Another notable point is that the loop functions of the diquarks can share the same structure with those of mesons. In fact, from Eq.~(\ref{Jud0+}) together with Eq.~(\ref{SaNdSC}), one can notice that the loop function of $[qq]_0^+$ is essentially the same as the pion's one.\footnote{The loop function of pion is given by
\begin{eqnarray}
{\cal J}_{\pi}= N_cT\sum_m\int\frac{d^3p}{(2\pi)^3}{\rm tr}\left[i\gamma_5 S_{(q)}(p')i\gamma_5 S_{(q)}(p)\right],
\end{eqnarray}
with $N_c=3$~\cite{Suenaga:2023tcy}. Regardless of the overall factor stemming from the color factor, indeed the structure is the same as ${\cal J}_{[qq]_0^+}$.} The duality is summarized as\footnote{This duality applies to the kernels (coupling constants) as well when the number of colors is $N_c=2$, i.e., in two-color QCD (QC$_2$D), owing to the so-called Pauli-G\"{u}rsey $SU(4)$ symmetry relating dynamics of mesons and diquarks~\cite{Kogut:1999iv,Kogut:2000ek}.} 
\begin{eqnarray}
0^+\  {\rm diquark }\ \ &\Leftrightarrow& \ \ 0^-\ {\rm meson} , \nonumber\\
0^-\ {\rm diquark }\ \ &\Leftrightarrow& \ \ 0^+\ {\rm meson} , \nonumber\\
1^+\ {\rm diquark }\ \ &\Leftrightarrow& \ \ 1^-\ {\rm meson} , \nonumber\\
1^-\ {\rm diquark }\ \ &\Leftrightarrow& \ \ 1^+\ {\rm meson} . \nonumber
\end{eqnarray}

Finally, we comment on the ``gauge invariance'' of the loop functions of (axial)vector diquarks. In association with the above duality between diquarks and mesons, the diquark one-loop functions should satisfy the Ward-Takahashi identities (WTIs) derived from the current (non)conservations. In fact, the loop structure of Eq.~(\ref{Jud1+}) together with charge-conjugation symmetry~(\ref{SaNdSC}) is found to be identical to the familiar vector current-current correlator as in quantum electrodynamics (QED)~\cite{peskin2018introduction}. When we employ the expressions of three-dimensional integrals as in the present work, the WTI for the spatial components are violated due to regularization artifacts at $T=0$. More concretely, one sees in Appendix~\ref{sec:LoopFunctions} that ${\cal J}_{\{qq\}_1^+}^{ij}|_{T=0}$ leads to quadratic divergences with a naive three-dimensional cutoff: ${\cal J}_{\{qq\}_1^+}^{ij}|_{T=0}\sim4\delta^{ij}/(3\pi^2)\Lambda_{\rm UV}^2 + \cdots$, while the other parts, ${\cal J}_{\{qq\}_1^+}^{00}$ and ${\cal J}_{\{qq\}_1^+}^{0j}$, do not. Here, we note that when the rest frame ${\bm q}={\bm 0}$ is adopted to estimate the diquark masses, the current (non)conservation law becomes trivial since in this limit there is no current flow. Thus, as long as we stick to the rest frame, at least the WTIs derived from the current (non)conservation are trivially satisfied despite the emergence of the quadratic divergences. Besides, the temperature-dependent parts of the loop functions which are the main target of this work are not suffered by the regularization artifacts, since these contributions are always finite due to the Boltzmann suppression factor. For these reasonings, in the present work, we proceed with the present three-dimensional-integral form. For more detail, see Appendix~\ref{sec:GaugeInvariance}. In this appendix, realization of the WTIs for $T$-dependent parts with general $q_0$ and ${\bm q}$ is explicitly shown in Eq.~(\ref{TDepWTI}).\footnote{One plausible way to evaluate the gauge invariant loop functions within the three-dimensional integral form is also provided in Appendix~\ref{sec:GaugeInvariantJs}.}

\begin{figure}[t]
\centering
\hspace*{-0.5cm} 
\includegraphics*[scale=0.45]{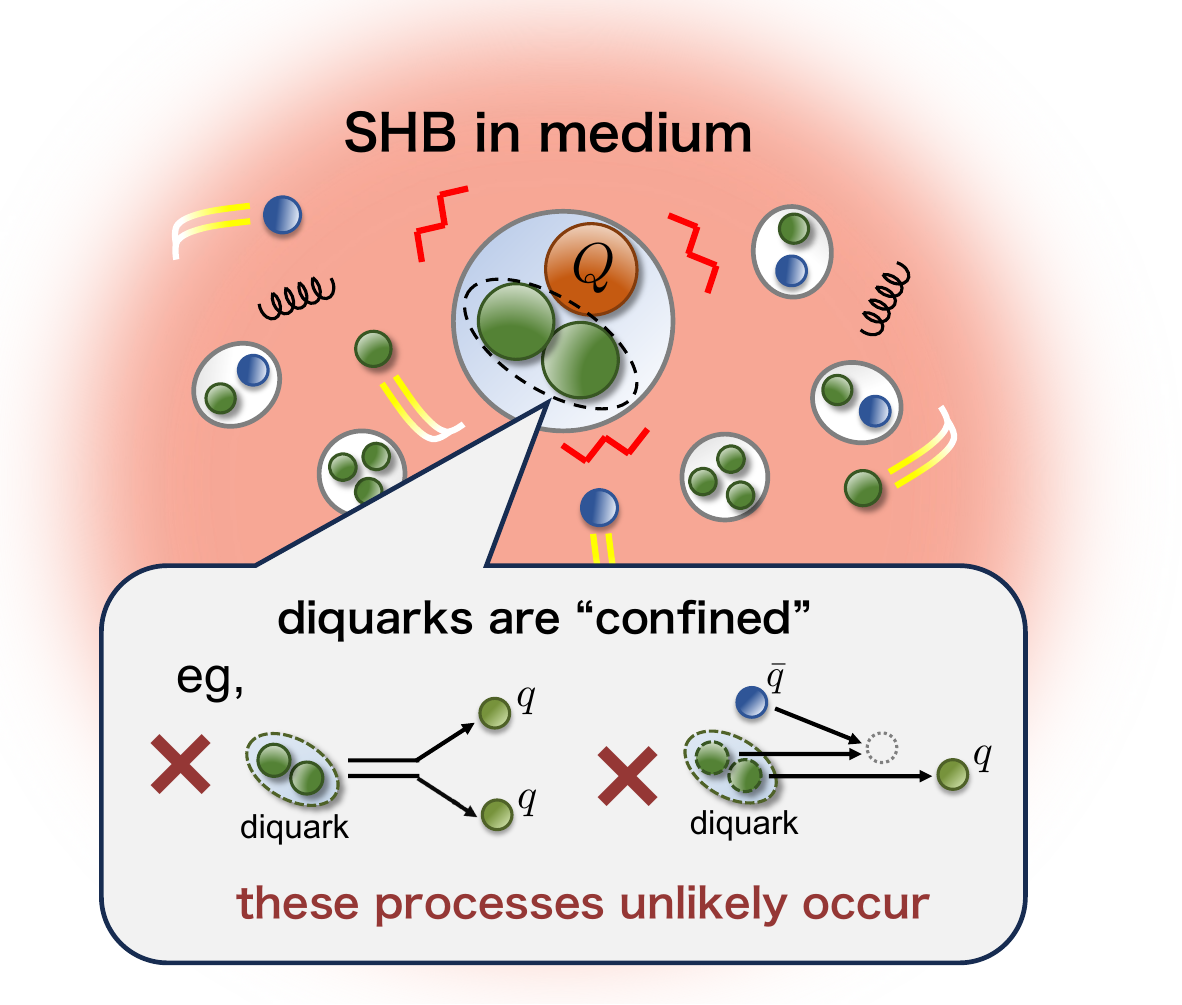}
\caption{A schematic picture that intuitively explains the absence of the imaginary parts.}
\label{fig:Schematic}
\end{figure}

\section{Regularization}
\label{sec:Regularization}

The UV divergences of the loop functions must be regularized when computing the scattering amplitudes numerically. In the present work we employ the three-dimensional proper-time regularization which includes both UV and infrared (IR) cutoffs, $\Lambda_{\rm UV}$ and $\mu_{\rm IR}$, as done in our previous work~\cite{Suenaga:2023tcy}.

 As inferred from Eq.~(\ref{Jud0+Comp}) or explicitly shown in Appendix.~\ref{sec:LoopFunctions}, the loop functions generally contain the $q_0$ dependences of $1/(q_0-\epsilon'_{{\bm p}'}-\epsilon_{\bm p}+i0)$, where we have included a infinitesimal positive imaginary part $+i0$ to manifestly represent the retarded propagation, due to the analytic continuation from the imaginary-time formalism. Our regularization is performed by replacing this piece as
\begin{eqnarray}
 \frac{1}{q_0-\epsilon'_{{\bm p}'}-\epsilon_{\bm p}+i0}  \to \frac{{\cal R}\left(q_0-\epsilon'_{{\bm p}'}-\epsilon_{\bm p}+i0 \right)}{q_0-\epsilon'_{{\bm p}'}-\epsilon_{\bm p}+i0},  \label{Replace}
\end{eqnarray}
where the regulator is of the form
\begin{eqnarray}
{\cal R}(x) \equiv {\rm e}^{-\frac{|x|}{\Lambda_{\rm UV}}} - {\rm e}^{-\frac{|x|}{\mu_{\rm IR}}} . \label{Regulator}
\end{eqnarray}
Equation~(\ref{Replace}) together with our proper-time regulator~(\ref{Regulator}) enables us to evaluate the imaginary part as 
\begin{eqnarray}
&& \frac{{\cal R}\left(q_0-\epsilon'_{{\bm p}'}-\epsilon_{\bm p}\right)}{q_0-\epsilon'_{{\bm p}'}-\epsilon_{\bm p}+i0} \nonumber\\
&&\ \ \  \overset{\rm Im}{\to} -i\pi{\cal R}\left(q_0-\epsilon'_{{\bm p}'}-\epsilon_{\bm p}\right)\delta\left(q_0-\epsilon'_{{\bm p}'}-\epsilon_{\bm p}\right). \label{RegulatorIm}
\end{eqnarray}
Here, the delta function is dealt with by the $p$ integral so that the right-hand side of Eq.~(\ref{RegulatorIm}) vanishes since ${\cal R}(0)=0$. In other words, our proper-time regularization allows us to eliminate the imaginary parts of the loop functions~\cite{Ebert:1996vx,Hellstern:1997nv}. Hence, one can easily define the mass of diquarks with no contamination from imaginary parts. Besides, this fact seems to be rather suitable to describe the diquarks confined in SHBs, since it is unlikely that those diquarks decay into quarks or are annihilated by antiquarks in medium due to a confining force from the remaining heavy quark. For a schematic picture see Fig.~\ref{fig:Schematic}. The IR cutoff would originate from nonperturbative gluodynamics at $\Lambda_{\rm QCD}$ of several hundreds MeV.

\begin{table}[htbp]
\begin{center}
  \begin{tabular}{c|cccc} \hline
Baryon & Diquark & $J^P$ & Mass & Total width  \\ \hline 
$\Lambda_c^+$ & $[ud]_0^+$ & $1/2^+$  & $2286.46$ & No strong decay  \\
$\Xi_c^{+}$ & $[su]_0^+$ & $1/2^+$  & $2467.71$ & No strong decay   \\
$\Xi_c^{0}$ & $[sd]_0^+$ & $1/2^+$ & $2470.44$ & No strong decay  \\
$\Sigma_c(2455)^{++}$ & $\{uu\}_1^+$ & $1/2^+$ & $2453.97$  & $1.89$  \\
$\Sigma_c(2455)^{+}$ & $\{ud\}_1^+$ & $1/2^+$  & $2452.65$ & $2.3$  \\
$\Sigma_c(2455)^{0}$ & $\{dd\}_1^+$ & $1/2^+$  & $2453.75$ & $1.83$  \\
$\Sigma_c(2520)^{++}$ & $\{uu\}_1^+$ & $3/2^+$ & $2518.41$ & $14.78$  \\
$\Sigma_c(2520)^{+}$ & $\{ud\}_1^+$ & $3/2^+$  & $2517.4$ & $17.2$  \\
$\Sigma_c(2520)^{0}$ & $\{dd\}_1^+$& $3/2^+$  & $2518.48$ & $15.3$  \\
$\Xi_c'^{+}$ & $\{su\}_1^+$ & $1/2^+$  & $2578.2$ & No strong decay   \\
$\Xi_c'^{0}$ & $\{sd\}_1^+$ & $1/2^+$  & $2578.7$ & No strong decay  \\
$\Xi_c(2645)^{+}$ & $\{su\}_1^+$ &  $3/2^+$  & $2645.10$ & $2.14$  \\
$\Xi_c(2645)^{0}$ & $\{sd\}_1^+$ & $3/2^+$ & $2646.16$ & $2.35$  \\
$\Omega_c^0$ & $\{ss\}_1^+$ & $1/2^+$  & $2696.2$ & $?$  \\
$\Omega_c(2770)^0$ & $\{ss\}_1^+$ & $3/2^+$ & $2765.9$ & $?$  \\
 \hline
 \end{tabular}
\caption{The PDG values (central values) of mass and width of singly charmed baryons in units of MeV~\cite{Workman:2022ynf}.}
\label{tab:MassWidth}
\end{center}
\end{table}

\begin{figure}[t]
\centering
\hspace*{-0.5cm} 
\includegraphics*[scale=0.6]{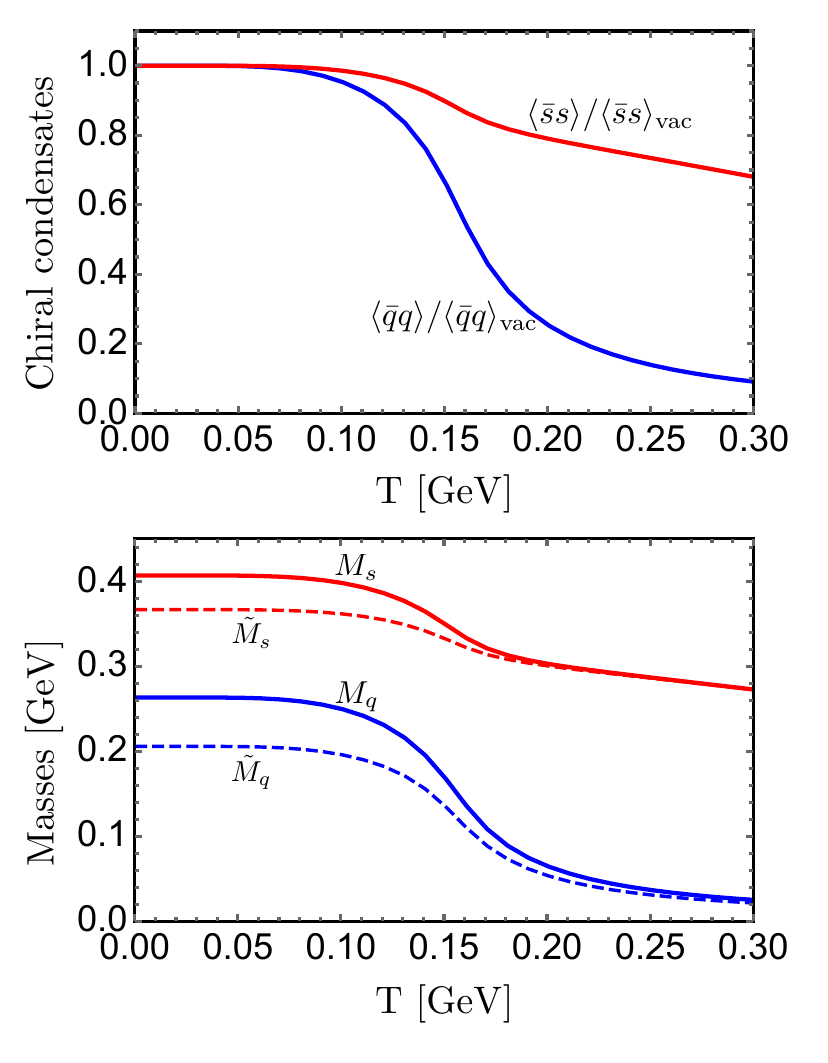}
\caption{Temperature dependences of the chiral condensates (top) and of the constituent quark masses (bottom).}
\label{fig:Condensate}
\end{figure}

\section{Numerical results}
\label{sec:Numerical}

In this section, we present numerical results on the masses and decay widths of the SHBs, more concretely the singly charmed baryons, at finite temperature, where the temperature effects are upon the diquarks inside the SHBs. Since the scalar and pseudoscalar diquraks carry spin $0$, the corresponding SHBs are HQS-singlet of $J^P=\frac{1}{2}^\pm$, with appropriate flavor structures. Meanwhile, the vector and axialvector ones carry spin $1$ so that they contribute to form the HQS-doublet of 
$J^P=\frac{1}{2}^\pm$ and $\frac{3}{2}^\pm$. We summarize experimental values of the masses and widths of the singly charmed baryons in the vacuum collected in the particle data group (PDG) in Table~\ref{tab:MassWidth}.

In what follows we use a notation of $\Lambda_c(\pm)$ and $\Xi_c(\pm)$ to refer to the HQS-singlet SHBs carrying $\pm$ parities. As for the HQS-doublet ones, we use $\Sigma_c(+)$, $\Xi_c'(+)$, $\Omega_c(+)$, $\Lambda'_c(-)$ and $\Xi_c'(-)$. The parity eigenvalues will be omitted when being trivially understood. There are two types of orbital excitations of the SHBs; one stems from the relative coordinate between a diquark and a heavy quark, and the other one from the excitation insides the diquark. In a quark-model language, the former and latter are called $\lambda$-mode and $\rho$-mode excitations, respectively, corresponding to the Jacobi coordinates generally adopted~\cite{Yoshida:2015tia}. Within our present approximation~(\ref{SHBMass}), the $\lambda$-mode excitations are ignored, and hence only the $\rho$-mode excitations made of the spin-$1$ diquarks~(\ref{VDQInt}) are examined. Those $\rho$-mode excitations are regarded as the chiral partners to the ground state $\Lambda_c(+)$ and $\Xi_c(+)$ within a chiral-model description. Experimentally observed $\{\Lambda_c(2595),\Lambda_c(2625)\}$ and $\{\Xi_c(2790),\Xi_c(2815)\}$ would be dominated by the $\lambda$-mode excitations since those SHBs are always lighter than the $\rho$-mode ones, whereas the $\rho$-mode ones, or the chiral-partner $\Lambda_c(-)$ and $\Xi_c(-)$, have not been observed.

\begin{table*}[htbp]
\begin{center}
\scalebox{1}{
  \begin{tabular}{c||cc|cccc|c} \hline
$K'$ [GeV$^{-5}$] & $H_S$ [GeV$^{-2}$] & $H_V$ [GeV$^{-2}$]  & ${\cal K}^{\rm vac}_{[qq]_0^+}$ [GeV$^{-2}$]  & ${\cal K}^{\rm vac}_{[qq]_0^-}$ [GeV$^{-2}$]  & ${\cal K}^{\rm vac}_{[sq]_0^+}$ [GeV$^{-2}$]  & ${\cal K}^{\rm vac}_{[sq]_0^-}$ [GeV$^{-2}$] & ${\cal K}_{1^\pm}$ [GeV$^{-2}$]  \\  \hline
$0$ & $2.20$ & $2.97$ & $4.40$  & $4.40$  & $4.40$  & $4.40$  & 5.94\\ \hline
$5$ & $2.08$ & $2.86$ & $4.31$ & $4.00$ & $4.27$ & $4.05$  & 5.72  \\ \hline
$10$ & $1.94$ & $2.73$ & $4.20$ & $3.56$ & $4.10$ & $3.66$ &  5.46  \\ \hline
$15$ & $1.76$ & $2.58$ & $3.99$ & $3.05$ & $3.85$ & $3.19$ & 5.16  \\ \hline
 \end{tabular}
 }
\caption{The estimated values of $H_S$ and $H_V$ and the corresponding vacuum kernels, with $K'[{\rm GeV}^{-5}]=0,5,10$ and $15$. The kernel for spin-$1$ is commonly given by ${\cal K}_{1^\pm} = 2H_V$ for all the channels. }
\label{tab:Kernel}
\end{center}
\end{table*}

\begin{table*}[htbp]
\begin{center}
\scalebox{1}{
  \begin{tabular}{c||ccccc|cccc} \hline
$K'$ [GeV$^{-5}$]  & $M^{\rm vac}_{[qq]_0^+}$ & $M^{\rm vac}_{[sq]_0^+}$ & $M^{\rm vac}_{\{qq\}_1^+}$ & $M^{\rm vac}_{\{sq\}_1^+}$ & $M^{\rm vac}_{\{ss\}_1^+}$  & $M^{\rm vac}_{[qq]_0^-}$ & $M^{\rm vac}_{[sq]_0^-}$ & $M^{\rm vac}_{[qq]_1^-}$  & $M^{\rm vac}_{[sq]_1^-}$  \\  \hline
$0$ & $0.197$ & $0.379$ & $0.412$ & $0.485$ & $0.534$ &  $0.425$ & $0.573$ & $0.636$ & $0.748$  \\ \hline
$5$ & $0.281$ & $0.463$ & $0.496$ & $0.560$ & $0.605$ & $0.619$ & $0.716$ & $0.706$ & $0.809$  \\ \hline
$10$ & $0.375$ & $0.558$ & $0.590$ & $0.639$ & $0.681$ &  $0.827$ & $0.877$ & $0.787$ & $0.881$  \\ \hline
$15$ & $0.502$ & $0.684$ & $0.717$ & $0.738$ & $0.766$ & $1.11$ & $1.10$ & $0.888$ & $0.972$  \\ \hline
 \end{tabular}
 }
\caption{The computed diquark masses in the vacuum for $K'[{\rm GeV}^{-5}]=0,5,10$ and $15$. The masses are displayed in a unit of GeV.}
\label{tab:VacuumMass}
\end{center}
\end{table*}

\subsection{Parameter determination}
\label{sec:Parameters}

Toward examination of temperature dependences of the SHB masses, here we determine the model parameters. 

Our NJL model contains nine parameters: $m_q$, $m_s$, $G_S$, $H_S$, $H_V$, $K$ and $K'$ from the Lagrangian~(\ref{LNJL}), and $\Lambda_{\rm UV}$ and $\mu_{\rm IR}$ for the cutoffs. Among these parameters, $m_q$, $m_s$, $G_S$, $K$, $\Lambda_{\rm UV}$ and $\mu_{\rm IR}$ are fixed solely by the meson sector and temperature dependences of the chiral condensates. As explained in Ref.~\cite{Suenaga:2023tcy} in detail, we choose the parameter values to be
\begin{eqnarray}
&& m_q=0.00258\, {\rm GeV}\ , \ \ m_s = 0.0761\, {\rm GeV}\ , \nonumber\\
&& G=1.15\, {\rm GeV}^{-2}\ , \ \ K=10.3\, {\rm GeV}^{-5}\ , \label{Parameters}
\end{eqnarray}
and
\begin{eqnarray}
\Lambda_{\rm UV} = 1.6\, {\rm GeV}\ , \ \ \mu_{\rm IR} = 0.45\, {\rm GeV}\ . \label{Cutoffs}
\end{eqnarray}
With these parameters, the decay constants $f_\pi=0.0921$ GeV, $f_K = 0.110$ GeV and the pseudoscalar-meson masses $m_\pi=0.138$ GeV, $m_K = 0.496$ GeV are reproduced. The cutoff values~(\ref{Cutoffs}) have been adjusted in such a way that the pseudocritical temperature with respect to chiral restoration reads $T_{\rm pc}\sim0.15$ GeV, as depicted in the top panel of Fig.~\ref{fig:Condensate}. This pseudocritical temperature is motivated by the lattice simulation results~\cite{Aoki:2009sc,HotQCD:2018pds}.

\begin{figure*}[t]
\centering
\hspace*{-0.5cm} 
\includegraphics*[scale=0.5]{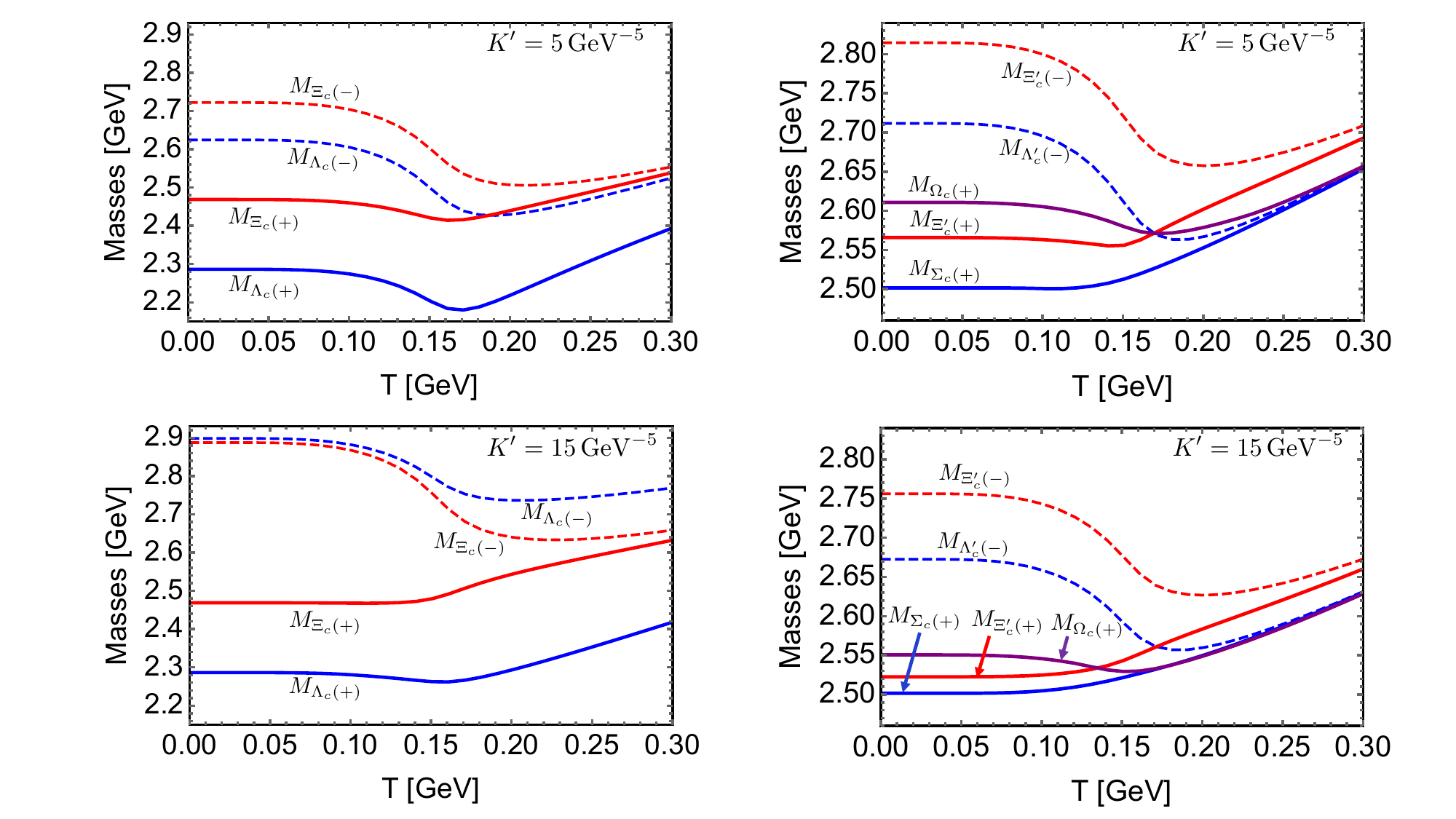}
\caption{ Temperature dependences of the SHB masses for $K'=5$ and $15$ GeV$^{-5}$.}
\label{fig:DiquarkMass}
\end{figure*}
\begin{figure*}[t]
\centering
\hspace*{-0.5cm} 
\includegraphics*[scale=0.5]{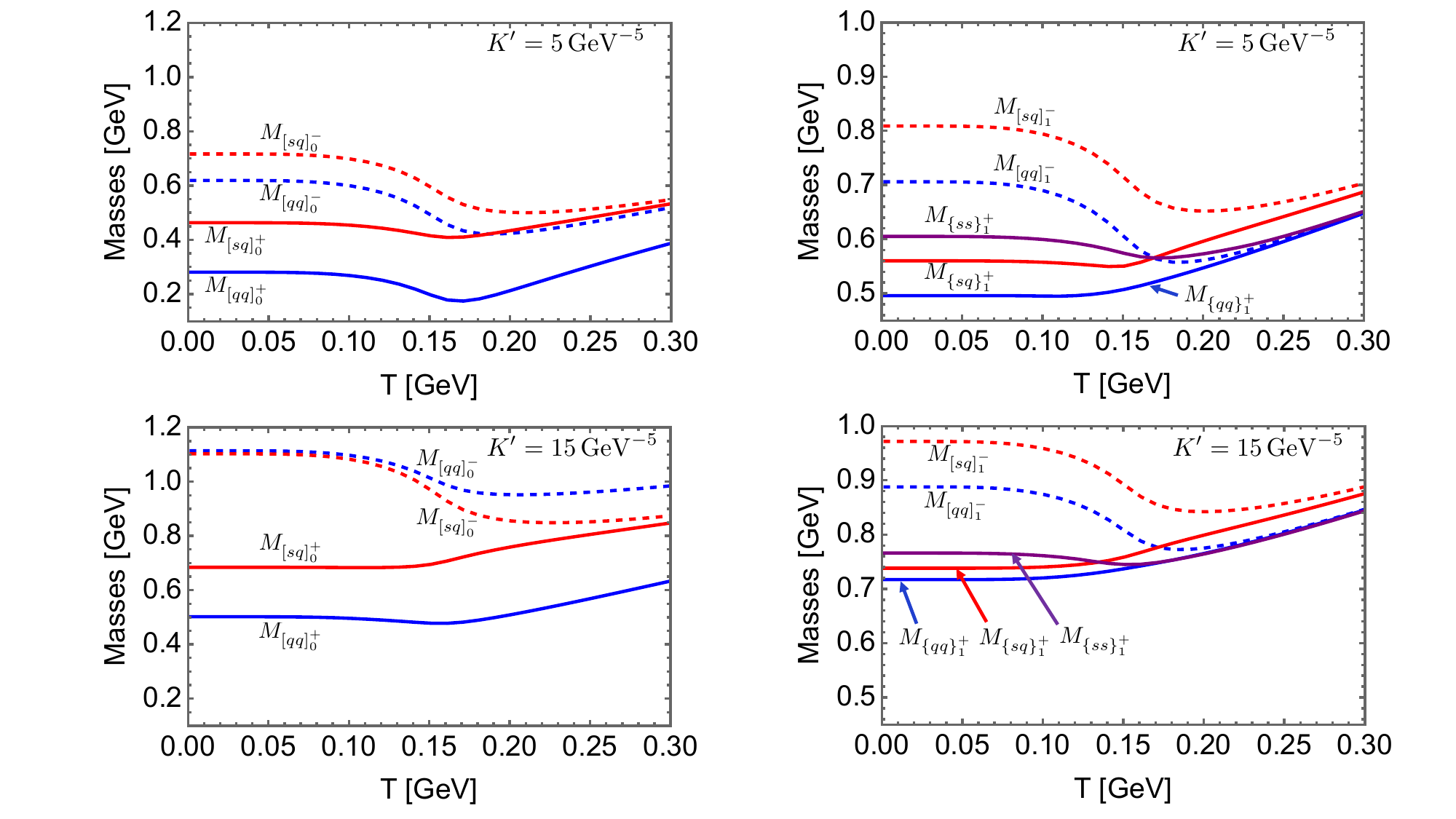}
\caption{ Temperature dependences of the spin-0 (left) and the spin-$1$ (right) diquark masses for $K'=5$ and $15$ GeV$^{-5}$.}
\label{fig:DiquarkLevel}
\end{figure*}

Temperature dependence of the constituent quark masses is also depicted in the bottom panel of Fig.~\ref{fig:Condensate}. The solid curves denote $M_q$ and $M_s$ defined in Eq.~(\ref{ConsMass}), which implies that the mass reductions of $M_q$ and $M_s$ follow the decrement of $\langle\bar{q}q\rangle$ and $\langle\bar{s}s\rangle$, respectively. This coincidence reflects a fact that $K$ term driving the flavor-mixing effects plays a minor role in the constituent quark masses. 
This behavior is more clearly seen by examining the $T$ dependences of $\tilde{M}_q$ and $\tilde{M}_s$ defined by subtracting $K$ terms in the masses:
\begin{eqnarray}
\tilde{M}_q &\equiv& M_q-2K\langle \bar{q}q\rangle\langle\bar{s}s\rangle , \nonumber\\
\tilde{M}_s &\equiv& M_q-2K\langle \bar{q}q\rangle^2 .
\end{eqnarray}
Indeed, the dashed curves in the bottom panel of Fig.~\ref{fig:Condensate} qualitatively trace the solid curves. The degeneracies of the $M_q$ and $\tilde{M}_q$ ($M_s$ and $\tilde{M}_s$) originate from the rapid decrement of $\langle\bar{q}q\rangle$ above $T_{\rm pc}$.


As for the remaining model parameters on the diquarks, $H_S$, $H_V$ and $K'$, we will keep $K'$ to be a free parameter so as to examine the $U(1)$ axial anomaly effects on the diquarks clearly. To fix $H_S$ and $H_V$ we adopt appropriate mass differences among the experimentally observed singly charmed baryons summarized in Table~\ref{tab:MassWidth}. In the present paper, the masses of SHBs, $M_{\rm SHB}$, are assumed to be determined by the following simple formula at any temperatures~\cite{Neubert:1993mb,manohar2000heavy}:
\begin{eqnarray}
M_{\rm SHB} = M_Q + M_{\rm diquark} , \label{SHBMass}
\end{eqnarray}
by virtue of the heavy-quark effective theory. In this formula $M_Q$ represents the mass of a constituent heavy quark which is universal to the arbitrary accompanying diquarks, while $M_{\rm diquark}$ is the diquark mass evaluated by the present NJL model. Hence, the mass differences are simply provided by the corresponding ones among the diquarks. For instance,
\begin{eqnarray}
&&\Delta m_1 \equiv  M^{\rm vac}_{\Xi_c} - M^{\rm vac}_{\Lambda_c} = M^{\rm vac}_{[sq]_0^+} -M^{\rm vac}_{[qq]_0^+} , \nonumber\\
&& \Delta m_2 \equiv  M^{\rm vac}_{\Sigma_c} - M^{\rm vac}_{\Lambda_c} = M^{\rm vac}_{\{qq\}_1^+} - M^{\rm vac}_{[qq]_0^+} , 
\end{eqnarray}
hold. Here $M^{\rm vac}_{\Lambda_c}$ and $M^{\rm vac}_{\Xi_c}$ are read off straightforwardly from Table~\ref{tab:MassWidth}. Meanwhile, since $\Sigma_c$'s are isotriplet and HQS-doublet of $J^P=\frac{1}{2}^+$ and $\frac{3}{2}^+$, we need to take isospin and spin averages to reasonably estimate $M^{\rm vac}_{\Sigma_c} $. The isospin average for spin-$1/2$ and spin-$3/2$ $\Sigma_c$ baryons are performed by
\begin{eqnarray}
M_{\Sigma_c(1/2^+)}^{\rm vac} &=& \frac{M^{\rm vac}_{\Sigma_c(2455)^{++}} + M^{\rm vac}_{\Sigma_c(2455)^{+}} + M^{\rm vac}_{\Sigma_c(2455)^{0}}}{3} , \nonumber\\
M_{\Sigma_c(3/2^+)}^{\rm vac} &=& \frac{M^{\rm vac}_{\Sigma_c(2520)^{++}} + M^{\rm vac}_{\Sigma_c(2520)^{+}} + M^{\rm vac}_{\Sigma_c(2520)^{0}}}{3} , \nonumber\\
\end{eqnarray}
respectively, and with these values the spin average yields
\begin{eqnarray}
M_{\Sigma_c}^{\rm vac} = \frac{M_{\Sigma_c(1/2^+)}^{\rm vac} + 3M_{\Sigma_c(3/2^+)}^{\rm vac}}{4} .
\end{eqnarray}
Thus, from Table~\ref{tab:MassWidth} $M_{\Sigma_c}^{\rm vac} = 2.50$ GeV. From the above estimation one can find
\begin{eqnarray}
\Delta m_1 = 0.183\, {\rm GeV}\ , \ \ \Delta m_2 = 0.215\, {\rm GeV} .\label{DeltaMInput}
\end{eqnarray}
In the following numerical analysis we will employ these two mass differences as further inputs to fix $H_S$ and $H_V$ for a given $K'$.

The numerically estimated coupling constants $H_S$ and $H_V$ and the corresponding vacuum kernels of the spin-$0$ diquarks for $K'\, [{\rm GeV}^{-5}]=0,5,10$ and $15$ are tabulated in Table~\ref{tab:Kernel}. This table indicates that the four-point interactions $H_S$ and $H_V$ get weakened as magnitude of the anomaly effect $K'$ is enhanced, and so do the all diquark kernels ${\cal K}$'s. That is, the attractive force between the quarks becomes weaker with the $U(1)$ axial anomaly effect increased. Accordingly, the diquarks become heavier for larger $K'$ as the binding energy gets small, which is clearly observed in Table~\ref{tab:VacuumMass}. For the scalar diquarks the masses are corrected by mixings with the time components of the vector-diquark loop function so that those naive structure would not hold easily. However, at least we have confirmed numerically that such mixing corrections are approximately universal for any choices of $K'$; the corrections commonly increase the diquark masses by $\sim 7$ \% for any $K'$. We note that the ratio of ${\cal K}^{\rm vac}_{[qq]_0^+}/{\cal K}^{\rm vac}_{[sq]_0^+}$ is also almost the same for all $K'$. Those universality would follow our inputs~(\ref{DeltaMInput}).

\begin{figure*}[t]
\centering
\hspace*{-0.5cm} 
\includegraphics*[scale=0.5]{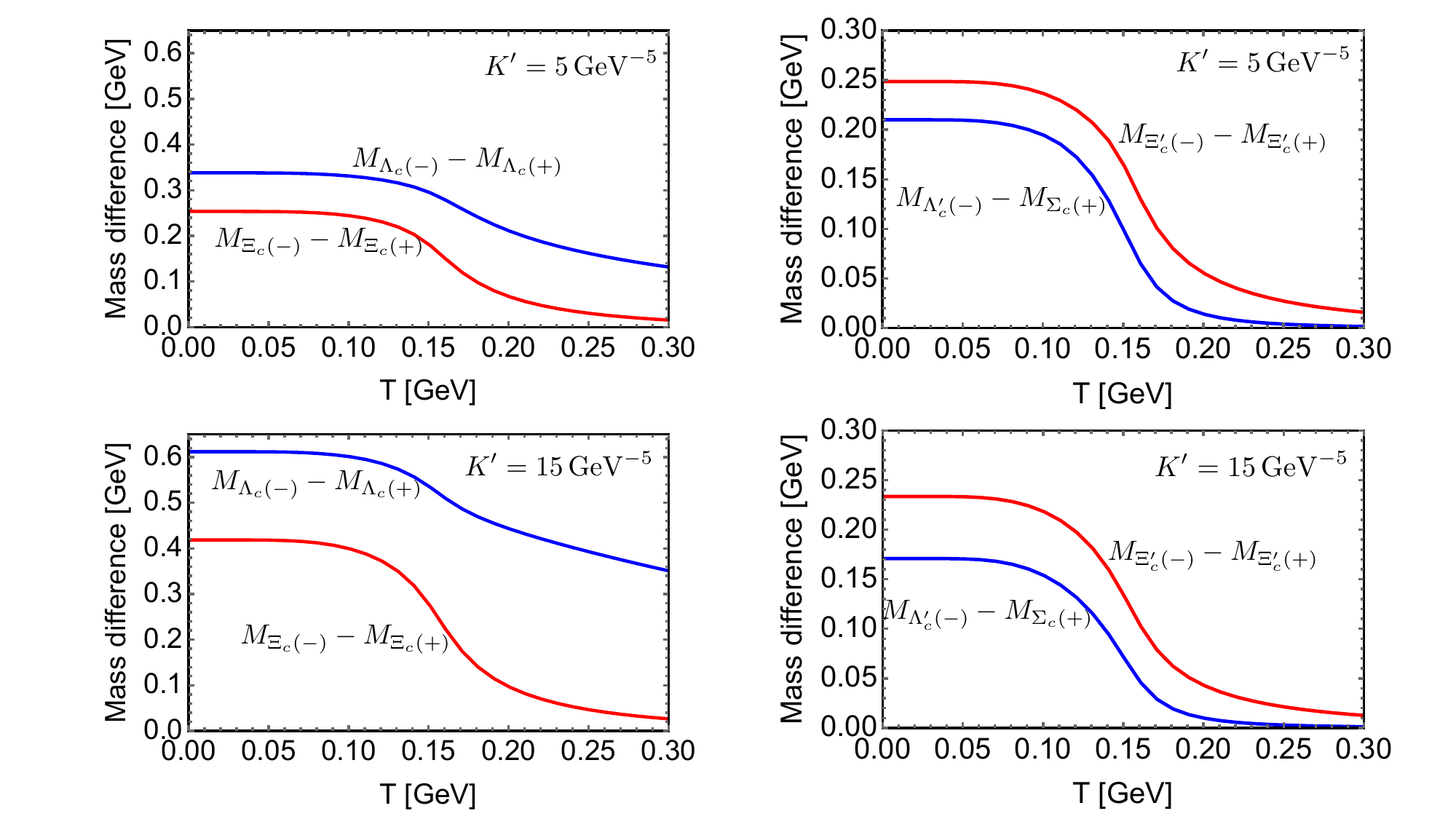}
\caption{ Temperature dependences of the mass differences of parity-partner SHBs for $K'=5$ and $15$ GeV$^{-5}$.}
\label{fig:DiquarkMassDiff}
\end{figure*}

\subsection{SHB masses at finite temperature}
\label{sec:SHBMass}

In this subsection, we present our main results on the fate of SHB masses at finite temperature.

Depicted in Fig.~\ref{fig:DiquarkMass} is the resultant $T$ dependences of the SHB masses for $K'=5$ GeV$^{-5}$ and $K'=15$ GeV$^{-5}$. The left panels are the masses of HQS-singlet SHBs formed by the (pseudo)scalar diquarks, while the right ones are those of HQS-doublet SHBs by the (axial)vector diquarks. From the left panel, one can see that the inverse mass ordering of the negative-parity and HQS-singlet SHBs, particularly at higher temperature, are realized for $K'=15$ GeV$^{-5}$; the $\Lambda_c(-)$ mass is heavier than the $\Xi_c(-)$ mass in contrast to our naive expectation from $M_s>M_q$~\cite{Harada:2019udr}. On the other hand, the positive-parity and HQS-singlet SHBs exhibit the naive mass ordering that is consistent with $M_s>M_q$. The figure also indicates that the masses, particularly $\Lambda_c$ mass, are once suppressed around $T\sim T_{\rm pc}$, which simply reflect the reduction of the constituent quark mass inside the diquarks stemming from the abrupt chiral restoration, as shown in Fig.~\ref{fig:Condensate}. Above this temperature thermal effects manifest themselves in addition to the chiral-restoration effects and govern the fate of the SHB masses, so that the SHB masses increase monotonically. The former mass reduction of $\Lambda_c$ is more prominent for $K'=5$ GeV$^{-5}$ since the $[qq]_0^+$ diquark mass is smaller and the reduction of $M_q$ affects considerably. Those characteristic temperature dependences of the masses of the HQS-singlet SHBs were indeed observed in our previous paper~\cite{Suenaga:2023tcy}.

It should be noted that within the present setup, the $[qq]_0^+$ diquark turns into a tachyonic mode around $T\sim T_{\rm pc}$ when a small value of $K'$ is taken, resulting in the unphysical instability of the ground state, and hence we have presented the numerical results with $K'\geq 5$ GeV$^{-5}$. For completeness we also depict temperature dependences of the spin-$0$ and spin-$1$ diquark masses in Fig.~\ref{fig:DiquarkLevel}, despite being essentially the same as Fig.~\ref{fig:DiquarkMass}.

The right panel of Fig.~\ref{fig:DiquarkMass} shows the masses of the HQS-doublet. It shows that the normal mass orderings consistent with the quark contents, follow at lower temperature for both the positive-parity and negative-parity SHBs, although quantitatively the masses of the positive-parity strange SHBs are underestimated. One reason to get this ordinary mass hierarchy is that the direct $U(1)$ axial anomaly effects on spin-$1$ diquarks leading to mixing different flavors are absent. Meanwhile, at higher temperature above $T_{\rm pc}\sim 0.15$ GeV, for the positive-parity SHBs, the mass of $\Omega_c$ decreases to degenerate with $\Sigma_c$. As a result, notably a new peculiar mass ordering of the HQS-doublet SHBs is predicted; $\Xi_c'$ gets heavier than $\Omega_c$ at high temperature. The spin-$1$ diquark masses evaluated within the NJL model will be analyzed in Sec.~\ref{sec:DiscussionDiquark} in detail.

Within the present treatment based on the approximation of Eq.~(\ref{SHBMass}), the binding energy between a diquark and a heavy quark cannot be estimated. For its evaluation, the diquark--heavy-quark potential model as adopted in Refs.~\cite{Kim:2020imk,Kim:2021ywp,Kim:2022pyq} is useful. Such improvement enables us to describe the dissociation of SHBs at high temperatures in a more realistic way. We leave these investigations for future study.

\subsection{Chiral-partner structures of the SHBs}
\label{sec:MassDifference}

The parity partners, hadrons carrying opposite parities, can be regarded as a useful indicator of chiral-symmetry restoration in medium, since their masses tend to be degenerated as chiral symmetry is restored to exhibit the so-called chiral-partner structure. To take a closer look at this characteristic phenomenon, we depict mass differences of the parity partners in Fig.~\ref{fig:DiquarkMassDiff} for $K'=5$ GeV$^{-5}$ and $K'=15$ GeV$^{-5}$. The left and right panels denote the mass differences among the HQS-singlet and HQS-doublet SHBs, respectively. Those figures indicate that the mass degeneracies between the partners are well realized except for the $\Lambda_c$ sector. This exceptional behavior stems from the slow reduction of $\langle\bar{s}s\rangle$ driven by the anomalous $K'$ term in the kernel~(\ref{Kernel0}), as analyzed in Ref.~\cite{Suenaga:2023tcy} in detail. It should be noted that the mass degeneracies of the HQS-doublet SHBs are always well realized since the $K'$ term does not contribute to them.

\section{Discussions on the spin-$1$ diquark mass spectrum}
\label{sec:DiscussionDiquark}

In Sec.~\ref{sec:SHBMass}, we have found characteristic mass hierarchies of the spin-$1$ diquarks; the small flavor dependence of the axialvector diquark masses, and the mass degeneracy of $\{qq\}_1^+$ and $\{ss\}_1^+$ diquarks at high temperature. Here, we discuss these issues focusing on structures of the loop functions analytically.

First, we explain the small flavor dependences of the axialvector diquarks, by comparing to those of the vector diquarks. Before proceeding with the arguments, we remind that the spin-$1$ diquark masses are evaluated by solving ${\cal J}(q_0) = {\cal K}^{-1}=({\rm constant})$ with respect to $q_0$, from Eq.~(\ref{BSEquation}). In other words, the quark mass dependence of the resultant diquark mass is obtained such that the loop function ${\cal J}$ is always constant.

For simplicity, here we restrict ourselves at zero temperature. In this limit, from the last equation in Eqs.~(\ref{JijAV}) and~(\ref{JVComp}), loop functions of the axialvector and vector diquarks generated by the identical quarks ($M_f\equiv M_1=M_2$) read
\begin{eqnarray}
{\cal J}_{\rm AV}^{ij} \to - 4\delta^{ij}\int\frac{d^3p}{(2\pi)^3}\frac{1}{\big(E_{\bm p}^{(f)}\big)^2}\left(\frac{4}{3}{\bm p}^2 + 2M_f^2\right) {\cal D}(q_0;{\bm p})\ ,  \label{JAVDemonstration} \nonumber\\
\end{eqnarray}
and
\begin{eqnarray}
{\cal J}_{\rm V}^{ij} \to - 4\delta^{ij}\int\frac{d^3p}{(2\pi)^3}\frac{1}{\big(E_{\bm p}^{(f)}\big)^2}\frac{4}{3}{\bm p}^2{\cal D}(q_0;{\bm p}) \label{JVDemonstration} \
\end{eqnarray}
at the rest frame ${\bm q}={\bm 0}$, respectively, with
\begin{eqnarray}
{\cal D}(q_0;{\bm p})\ \equiv \frac{1}{q_0-2E_{{\bm p}}^{(f)}} -\frac{1}{q_0+2E_{{\bm p}}^{(f)} }\ . \label{DDef}
\end{eqnarray} 
The difference between axialvector and vector diquark channels are the sign of quark mass contributions, in such a way as to leave $2M_f^2$ only for ${\cal J}_{\rm AV}$. Here, the loop functions are governed by the excitations of ${\bm p}\sim 1$ GeV for which the contributions~(\ref{DDef}) are mostly negative. Hence, in order to make the loop function constant, when we change the quark mass $M_f$, a smaller variation of $q_0$ is enough for ${\cal J}_{\rm AV}$ compared to ${\cal J}_{\rm V}$ due to the compensation from $2M_f$ in Eq.~(\ref{JAVDemonstration}). As an extreme situation, when we ignore the detail of regularization and assume the integrands of Eqs.~(\ref{JAVDemonstration}) and~(\ref{JVDemonstration}) themselves are constant, the following conditions are obtained:
\begin{eqnarray}
\frac{\frac{4}{3}{\bm p}^2+2M_f^2}{q_0^2-4\big(E_{\bm p}^{(f)}\big)^2} \sim -{\cal C}_{\rm AV}\  , \ \ \frac{\frac{4}{3}{\bm p}^2}{q_0^2-4\big(E_{\bm p}^{(f)}\big)^2} \sim -{\cal C}_{\rm V}\ ,
\end{eqnarray}
where ${\cal C}_{AV},{\cal C}_V>0$ are constants, for a given ${\bm p}$ ($\sim1$ GeV). These conditions yield
\begin{eqnarray}
q_0^2 \sim \left(4-\frac{2}{{\cal C}_{\rm AV}}\right)M_f^2 + \cdots\ , \ \ q_0^2 \sim 4M_f^2 + \cdots\ , 
\end{eqnarray}
for the axialvector and vector channels, respectively.
Therefore, a small correction of $q_0$ with respect to a change of $M_f$ is indeed obtained for the axialvector diquarks. Those structures essentially lead to the smaller flavor dependence for the axialvector diquarks, as indicated in the right panel of Fig.~\ref{fig:DiquarkLevel}, which would partly explain the small mass splittings among the axialvector diquarks.

We note that a similar argument follows for the spin-$0$ diquarks apart from the anomaly effects and mixings from the spin-$1$ diquarks; From Eqs.~(\ref{JSApp}) and~(\ref{JPSApp}) one can expect that the flavor dependence is smaller for the scalar diquarks compared to the pseudoscalar ones, and in fact these tendencies are reflected in our previous simplified analysis~\cite{Suenaga:2023tcy}. The present computations are essentially the same as those of the mesons as long as the one-loop analysis within the NJL model is adopted. So we also note that a similar small flavor dependence for the vector meson masses is obtained, as done in, e.g., Ref.~\cite{Takizawa:1990ku} where those masses are underestimated.

Next, we explain the mass degeneracy of $\{qq\}_1^+$ and $\{ss\}_1^+$ diquarks at higher temperature exhibited in the right panel of Fig.~\ref{fig:DiquarkLevel}. From Eq.~(\ref{JijAV}) the loop functions of $\{qq\}_1^+$ and $\{ss\}_1^+$ diquark channels share the common piece of $1-2f_F(E_{\bm p}^{(f)})$ ($f=q,s$) in their integrands. Those contributions are mostly unity for all range of ${\bm p}$ due to the rapid suppression of the distribution function, as long as temperature is small compared to $M_f$. Hence, soft modes which are sensitive to the mass difference of $M_q$ and $M_s$ survive sizably so as to generate the mass difference of $\{qq\}_1^+$ and $\{ss\}_1^+$ in such a cold system. Meanwhile, the factor $1-2f_F(E_{\bm p}^{(f)})$ is largely suppressed for the soft modes, when temperature is adequately high, and thus the variation from $M_q$ and $M_s$ is diminished. This suppression results in the universality of the loop functions and the masses of $\{qq\}_1^+$ and $\{ss\}_1^+$ get degenerated. On the other hand, the loop function of $\{sq\}_1^+$ includes additional contributions proportional to $f_F(E_{\bm p}^{(q)})-f_F(E_{\bm p}^{(s)})$ from Eq.~(\ref{JijAV}). Therefore, the mass of $\{sq\}_1^+$ ($\Xi_c'$) always deviates from those of $\{qq\}_1^+$ and $\{ss\}_1^+$ ($\Sigma_c$ and $\Omega_c$) even at higher temperature.

\section{Decay width of $\Sigma_c$ at finite temperature}
\label{sec:DecaySigmaC}

In Sec.~\ref{sec:Numerical}, mass modifications of the HQS-singlet and HQS-doublet SHBs at finite temperature are comprehensively predicted based on a chiral model for the corresponding diquarks. In particular, the mass degeneracies of the parity partners driven by the chiral restoration are clearly delineated. Here, we focus on the fate of decays of $\Sigma_c\to\Lambda_c\pi$ to present a clear signal of the predicted mass spectra.

\subsection{Decay formula}
\label{sec:DecayFormula}

An interaction Lagrangian which is capable of describing decays of $\Sigma_c\to\Lambda_c\pi$ is given by~\cite{Suenaga:2022ajn}
\begin{eqnarray}
{\cal L}_{\rm int} = G_A{\rm tr}\left[\bar{\cal B}_R\partial_\mu\Sigma^\dagger {\cal B}_{\bm 6}^{T,\mu} + \bar{\cal B}_L\partial_\mu\Sigma {\cal B}_{\bm 6}^\mu + {\rm h.c.}\right] ,\label{DecayLagrangian}
\end{eqnarray}
where $({\cal B}_{L(R)})_{ij} \sim  \epsilon_{ijk}Q^a(\eta_{L(R)})_k^a$ denotes the HQS-singlet SHBs while $({\cal B}^{\mu}_{{\bm 6}})_{ij}\sim Q^a(\tilde{\eta}_{{\bm 6}})^{a,\mu}_{ij}$ denotes the HQS-doublet ones. More concretely
\begin{eqnarray}
{\cal B}_\pm &=& \frac{1}{\sqrt{2}}({\cal B}_R\mp {\cal B}_L) \nonumber\\
&=& \left(
\begin{array}{ccc}
0 & \Lambda_c(\pm) & -\Xi_c(\pm) \\
-\Lambda_c(\pm) & 0 & \Xi_c(\pm) \\
\Xi_c(\pm) & -\Xi_c(\pm) & 0 \\
\end{array}
\right) , \label{HQS-S}
\end{eqnarray}
and 
\begin{eqnarray}
{\cal B}_{\bm 6} = \left(
\begin{array}{ccc}
\Sigma_c^{I_z=1} & \frac{1}{\sqrt{2}}\Sigma_c^{I_z=0} & \frac{1}{\sqrt{2}}\Xi_c'^{I_z=1/2} \\
\frac{1}{\sqrt{2}}\Sigma_c^{I_z=0} & \Sigma_c^{I_z=-1} & \frac{1}{\sqrt{2}}\Xi_c'^{I_z=-1/2} \\
\frac{1}{\sqrt{2}}\Xi_c'^{I_z=1/2} & \frac{1}{\sqrt{2}}\Xi_c'^{I_z=-1/2} & \Omega_c \\
\end{array}
\right) , \label{HQS-D}
\end{eqnarray}
where $I_z$ in Eq.~(\ref{HQS-D}) stands for the isospin third component. Another $3\times3$ matrix $\Sigma$ in Eq.~(\ref{DecayLagrangian}) is a scalar and pseudoscalar meson nonet  $\Sigma= S+iP$ in which the latter is embedded as
\begin{eqnarray}
P  = \frac{1}{\sqrt{2}}\left(
\begin{array}{ccc}
\frac{\eta_N+\pi^0}{\sqrt{2}} & \pi^+& K^+ \\
\pi^- & \frac{\eta_N-\pi^0}{\sqrt{2}}  & K^0 \\
K^-& \bar{K}^0 & \eta_S \\
\end{array} 
\right) .
\end{eqnarray}
This nonet transforms under $SU(3)_L\times SU(3)_R$ as $\Sigma\to g_L\Sigma g_R^\dagger$. Hence, from the transformation laws~(\ref{FieldTrans}) with the decomposition~(\ref{EtaTildeDec}) for the SHBs, chiral symmetry of the interactions in Eq.~(\ref{DecayLagrangian}) is obvious.\footnote{From the interpolating field~(\ref{OperatorsChiral}), the chiral transformation law of ${\cal B}_{L(R)}$ is understood to be ${\cal B}_{L(R)}\to g_{L(R)}{\cal B}_{L(R)}g^T_{L(R)}$.} Moreover, this interaction Lagrangian is invariant under $SU(2)$ HQS transformation since no gamma matrices driving spin flippings of the heavy quarks are present.

\begin{figure*}[t]
\centering
\hspace*{-0.5cm} 
\includegraphics*[scale=0.5]{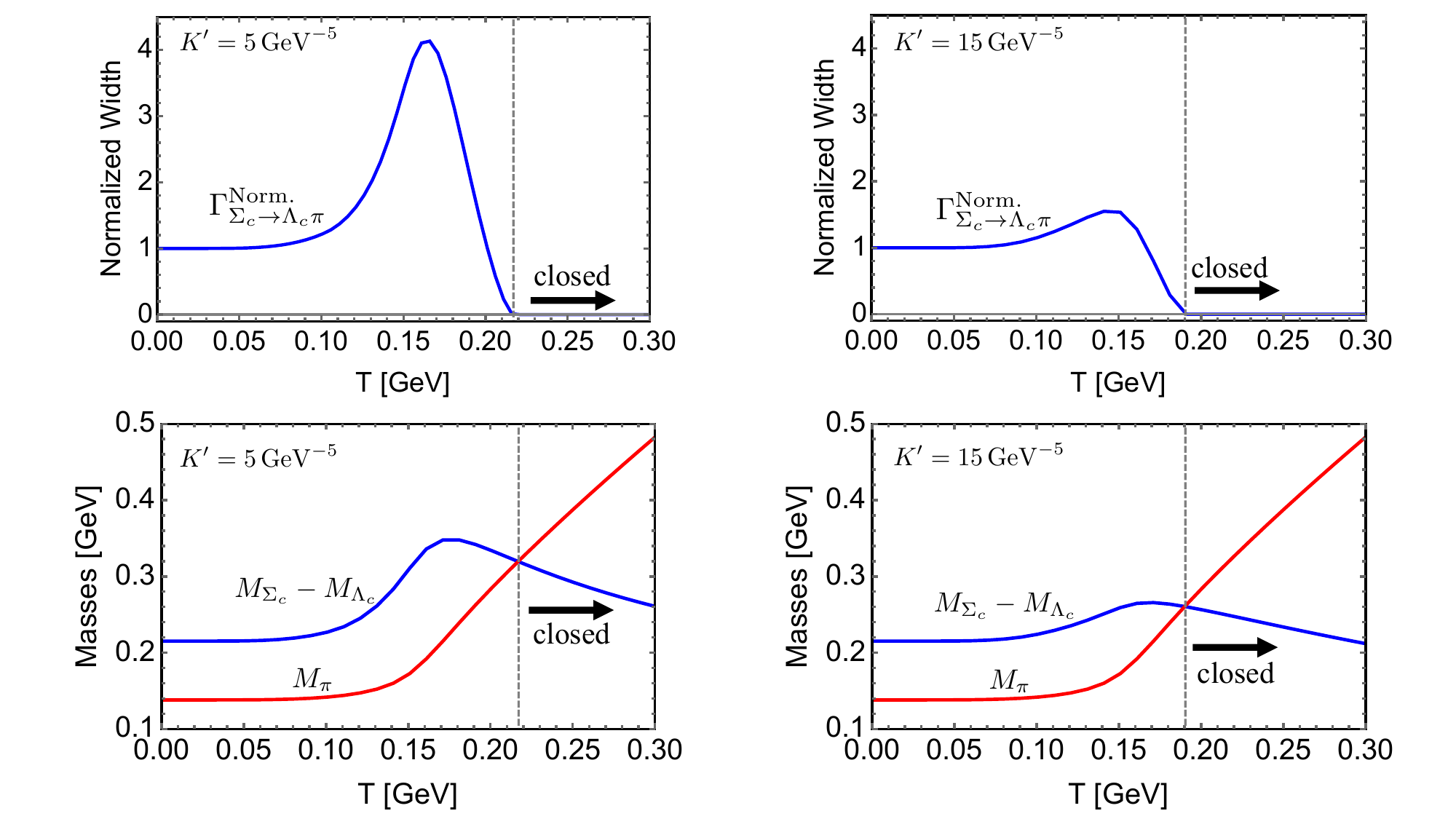}
\caption{Top: temperature dependences of the normalized decay width $\Gamma^{\rm Norm.}_{\Sigma_c\to\Lambda_c\pi}$ for $K'=5$ GeV$^{-5}$ and $K'=15$ GeV$^{-5}$ defined in Eq.~(\ref{GammaNorm}). Bottom: temperature dependences of the mass difference $M_{\Sigma_c}-M_{\Lambda_c}$ and the pion mass $M_\pi$. }
\label{fig:NormWidth}
\end{figure*}

From the Lagrangian~(\ref{DecayLagrangian}), the interactions corresponding to $\Sigma_c\to\Lambda_c\pi$ are read off to be
\begin{eqnarray}
{\cal L}_{\rm int} &\to& iG_A\Big[\bar{\Lambda}_c\partial_\mu\pi^-(\Sigma_c^{I_z=1})^\mu-\bar{\Lambda}_c\partial_\mu\pi^0(\Sigma_c^{I_z=0})^\mu  \nonumber\\
&& - \bar{\Lambda}_c\partial_\mu\pi^+(\Sigma_c^{I_z=-1})^\mu\Big] .
\end{eqnarray}
Here, the HQS-doublet field $(\Sigma^{I_z}_c)^\mu$ contains a spin-$1/2$ and spin-$3/2$ components, $\Sigma_c^{I_z}$ and $(\Sigma_c^{I_z})^{*\mu}$. By separating these components as 
\begin{eqnarray}
(\Sigma_c^{I_z})^\mu = (\Sigma_c^{I_z})^{*\mu} + \frac{1}{\sqrt{3}}(\gamma^\mu + v^\mu)\gamma_5 \Sigma_c^{I_z} ,
\end{eqnarray}
where $v^{\mu}$ is the four-velocity of the baryon, the decay widths of $\Sigma_c\to\Lambda_c\pi$ are evaluated. The resultant decay width reads~\cite{Suenaga:2022ajn}
\begin{eqnarray}
\Gamma_{\Sigma_c\to\Lambda_c\pi} = \frac{G_A^2}{12\pi}\frac{(E_{\Lambda_c}+M_{\Lambda_c})|{\bm p}_\pi|^3}{M_{\Sigma_c}} , \label{SigmaDecay}
\end{eqnarray}
for all the channels, where $E_{\Lambda_c} = \sqrt{M_{\Lambda_c}^2 + |{\bm p}_\pi|^2}$ is the dispersion relation of the final-state $\Lambda_c$, and ${\bm p}_\pi$ denotes the momentum of the emitted pion that is simply given by combinations of the hadron masses from energy-momentum conservations. Assuming that the coupling manners and dispersion relations do not change in medium significantly, one can adopt Eq.~(\ref{SigmaDecay}) at any temperatures. 
 In particular, corrections of the decay width of $\Sigma_c\to\Lambda_c\pi$ at finite temperature are examined by the following normalized decay width
 \begin{eqnarray}
\Gamma^{\rm Norm.}_{\Sigma_c\to\Lambda_c\pi} \equiv \frac{\Gamma_{\Sigma_c\to\Lambda_c\pi}}{\Gamma^{\rm vac}_{\Sigma_c\to\Lambda_c\pi}} = \frac{(E_{\Lambda_c}+M_{\Lambda_c})|{\bm p}_\pi|^3}{(E^{\rm vac}_{\Lambda_c}+M^{\rm vac}_{\Lambda_c})|{\bm p}^{\rm vac}_\pi|^3}\frac{M^{\rm vac}_{\Sigma_c}}{ M_{\Sigma_c}} , \nonumber\\ \label{GammaNorm}
\end{eqnarray} 
with no ambiguities in determining $G_A$.

\subsection{$\Gamma^{\rm Norm.}_{\Sigma_c\to\Lambda_c\pi}$ at high temperature}
\label{sec:DecayResult}

In this subsection, we examine the normalized decay width $\Gamma^{\rm Norm.}_{\Sigma_c\to\Lambda_c\pi}$ in Eq.~(\ref{GammaNorm}) at finite temperature. 

Depicted in the top panels of Fig.~\ref{fig:NormWidth} denotes the resultant $T$ dependences of $\Gamma^{\rm Norm.}_{\Sigma_c\to\Lambda_c\pi}$ for $K'=5$ GeV$^{-5}$ and $K'=15$ GeV$^{-5}$. From these figures one can see that the decay width is once enhanced around $T\sim T_{\rm pc}$, and then it falls to zero at $T\approx0.22$ GeV and $T\approx0.19$ GeV for $K'=5$ GeV$^{-5}$ and $K'=15$ GeV$^{-5}$, respectively. That is, it is shown that the decay width of $\Sigma_c\to\Lambda_c\pi$ vanishes above the pseudocritical temperature $T_{\rm pc}$. The enhancement at $T\sim T_{\rm pc}$ is more prominent for smaller $K'$.

In order to investigate this vanishing decay width in detail, we also plot the temperature dependences of the mass difference $M_{\Sigma_c}-M_{\Lambda_c}$ and the pion mass $M_\pi$ in the bottom panels of Fig.~\ref{fig:NormWidth}.\footnote{The pion mass $M_\pi$ at finite temperature is determined in a similar way to the diquark masses. For more detail see Ref.~\cite{Suenaga:2023tcy}.} The mass difference $M_{\Sigma_c}-M_{\Lambda_c}$ gets enhanced around $T\sim T_{\rm pc}$ reflecting the reduction of $M_{\Lambda_c}$ as displayed in the left panels of Fig.~\ref{fig:DiquarkMass}. After this enhancement the mass difference monotonically decreases due to the rather rapid increase of $M_{\Lambda_c}$ at high-temperature regime. Meanwhile the pion mass $M_\pi$ monotonically increases especially above $T_{\rm pc}$ from thermal effects. As a result, $M_{\Sigma_c}-M_{\Lambda_c}<M_\pi$ is satisfied, i.e., the threshold is closed at a certain temperature, leading to the vanishing of the decay width. We note that the enhancement of $M_{\Sigma_c}-M_{\Lambda_c}$ around $T\sim T_{\rm pc}$ is more prominent for smaller values of $K'$, since the smaller $K'$ we take the larger reduction of $M_{\Lambda_c}$ we obtain, as discussed in detail in Sec.~\ref{sec:SHBMass}. This is also a reason why the threshold closing takes place at higher temperature for the smaller $K'$.

As tabulated in Table~\ref{tab:MassWidth} the $\Xi_c(2645)^+$ and $\Xi_c(2645)^0$ baryons can possess finite decay widths due to $\Xi_c'\to\Xi_c\pi$. In our present analysis unfortunately this decay width cannot be reproduced since we have treated spin-$1/2$ and spin-$3/2$ $\Xi_c'$ baryons collectively by assuming the HQSS, and moreover their masses have been estimated to be $2.57$ GeV in the vacuum, which leads to vanishing of the $\Xi_c'\to\Xi_c\pi$ decay. However, still one can infer the fate of $\Xi_c'\to\Xi_c\pi$ decay width at high temperature by seeing the mass difference of $M_{\Xi_c'}-M_{\Xi_c}$. This mass difference does not grow with increasing temperature, while the pion mass $M_\pi$ monotonically evolves due to the thermal effects above $T\sim T_{\rm pc}$. Hence, the decay widths of $\Xi_c(2645)^+$ and $\Xi_c(2645)^0$ driven by pion emissions are also expected to be closed above $T_{\rm pc}$ similarly to the widths of $\Sigma_c$ baryons.

In the present analysis we have presented the results with the averaged masses of $\Sigma_c$ and $\Sigma_c^*$ based on the HQSS. Even when we take into account its violation to estimate their masses more precisely, the closing of the decay modes is expected to occur for all the channels.

\section{Conclusions}
\label{sec:Conclusion}

In this paper we have examined the mass corrections of the HQS-singlet and HQS-doublet singly charmed baryons at finite temperature from HQSS and the restoration of chiral symmetry. Modifications and chiral dynamics of the SHBs in medium are incorporated through the composing diquarks by virtue of the heavy-quark effective theory.

The diquarks are described by the NJL model which includes (pseudo)scalar-type and (axial)vector-type four-point interactions. Besides, the six-point interaction violating only $U(1)$ axial symmetry is adopted to provide the $U(1)$ axial anomaly effects on the spin-$0$ diquarks. The divergences are handled by means of the three-dimensional proper-time regularization including both the UV and IR cutoffs, so as to eliminate imaginary parts which correspond to unphysical processes for confined diquarks inside SHBs.

The resultant mass spectrum of the HQS-singlet SHBs shows that the inverse mass hierarchy where the mass of $\Lambda_c(-)$ becomes larger than that of $\Xi_c(-)$ despite their quark contents~\cite{Harada:2019udr}, is realized with more significant anomaly effects, especially at high temperature regime. These behaviors are consistent with the predictions in our previous work~\cite{Suenaga:2023tcy} where different inputs were adopted. For HQS-doublet SHBs carrying positive parities, a mass degeneracy of $\Omega_c$ and $\Sigma_c$ has been predicted above the pseudocritical temperature $T_{\rm pc}$ due to the common thermal effects for those baryons. This remarkable mass degeneracy leads to a new peculiar mass ordering where $\Xi'_c$ gets heavier than $\Omega_c$.

The mass degeneracies of the parity partners, i.e., the chiral-partner structures, have been predicted for all the SHBs. For HQS-singlet SHBs, the mass degeneracies of $\Lambda_c$ sector are found to be delayed when the anomaly effects are sufficient, reflecting the slow reduction of $\langle\bar{s}s\rangle$. The mass degeneracies of the HQS-doublet SHBs are always well realized reflecting the absence of the anomaly effects. 

In addition, we have found that the decay width of $\Sigma_c\to\Lambda_c\pi$ is once enhanced around the pseudocritical temperature $T_{\rm pc}$ and then converges to zero above $T_{\rm pc}$. Similarly we predict that the decay channel of $\Xi_c'\to\Xi_c\pi$ also gets closed at high temperature.

The predicted masses and fate of the decay widths are expected to provide future heavy-ion collision experiments and lattice simulations with useful information on the chiral-symmetry properties of the spin-$0$ and spin-$1$ diquarks.

In what follows we give some comments on the present work. In this paper we have employed the three-dimensional integral form with the proper-time regularization to regularize the one-loop functions. As explained in Sec~\ref{sec:BSEquation} this scheme can barely respect the WTIs derived from appropriate current (non)conservations in defining the masses of the spin-$1$ diquarks, since at the rest frame the currents do not flow and the identities trivially hold. On the other hand, the WTIs are violated when we extend the analysis to the finite momentum system due to the current flow for the spatial components of the spin-$1$ diquark. Thus, it is inevitable to improve this issue to generalize the examination for the moving diquarks.

In three-color QCD that is corresponding to our real-life world, the diquarks are not direct observables since they are colorful, but the SHBs can be regarded as alternative color-single observables. Meanwhile, in two-color QCD, QC$_2$D, the diquarks are counted as color-singlet baryons and their intrinsic properties such as the masses are defined in a proper way. Indeed, the lattice simulations have been reporting numerical results on the hadron mass spectrum including the diquark baryons~\cite{Hands:2007uc,Murakami:2022lmq}. In other words, QC$_2$D world has a great advantage in pursuing the diquark dynamics. In fact, although in three-color QCD the $\Lambda_c(-)$ has not been experimentally observed, on the QC$_2$D lattice the mass of $0^-$ diquark was indeed computed. In this regard, numerical studies in QC$_2$D together with theoretical analyses based on chiral models~\cite{Kogut:1999iv,Ratti:2004ra,Suenaga:2022uqn,Suenaga:2023xwa} would also be one of the promising approaches toward understanding the diquarks and SHBs.

Our present NJL model analysis naturally predicts $\Lambda_c(-)$ and $\Xi_c(-)$ as the chiral partners to the ground-state $\Lambda_c$ and $\Xi_c$, which play a key role in describing the SHBs from the linear representation of chiral symmetry, despite not being experimentally observed. For instance, the main decay mode of $\Lambda_c(-)$ is $\Lambda_c(-) \to \Lambda_c(+)\eta$, the threshold of which is $2834$ MeV in the vacuum. Meanwhile, as exhibited in Fig.~\ref{fig:DiquarkMass} the predicted mass of $\Lambda_c(-)$ is largely dependent on the magnitude of the anomaly effect $K'$. In particular, when the value of $K'$ is sufficiently large, the decay width reads several hundred of MeV which is hopeless to observed $\Lambda_c(-)$ due to its broad phase space. When we access finite temperature, on the other hand, the mass difference between $\Lambda_c(+)$ and $\Lambda_c(-)$ gets small, resulting in the suppression of the decay width, as analyzed in Ref.~\cite{Suenaga:2023tcy} in detail. Hence, it would be possible to confirm the existence of $\Lambda_c(-)$ in such hot medium. We leave more phenomenological analyses on this issue, together with the fate of decays of the HQS-doublet SHBs, for future study.

\section*{Acknowledgement}

This work was supported by Grants-in-Aid for Scientific Research No.~23K03377 and No.~23H05439 (D.S.), and No.~23K03427 (M.O.), from Japan Society for the Promotion of Science.

\appendix

\section{One-loop functions}
\label{sec:LoopFunctions}

In this appendix, we present one-loop functions of the diquarks. Here, in order to keep rather general discussions, we assume that the loop functions are generated by two quarks whose masses are $M_1$ and $M_2$ where vertex structures are appropriately provided.

We defined loop functions of the scalar-diquark, pseudoscalar-diquark, axialvector-diquark, and vector-diquark channels by ${\cal J}_{\rm S}$, ${\cal J}_{\rm PS}$, ${\cal J}_{\rm AV}$, and ${\cal J}_{\rm V}$, respectively. Those with two different quarks are of the forms
\begin{eqnarray}
&& {\cal J}_{\rm S} = -2T\sum_m\int\frac{d^3p}{(2\pi)^3}{\rm tr}\Big[\gamma_5S_1(p')\gamma_5S_2^{\cal C}(p) +(1\leftrightarrow 2) \Big] ,\nonumber\\
&& {\cal J}_{\rm PS} = 2T\sum_m\int\frac{d^3p}{(2\pi)^3}{\rm tr}\Big[S_1(p')S_2^{\cal C}(p) + (1\leftrightarrow 2)\Big] , \nonumber\\
&& J^{\mu\nu}_{\rm AV} = 2T\sum_m\int\frac{d^3p}{(2\pi)^3} {\rm tr}\Big[ \gamma^\mu S_1(p')\gamma^\nu S_2^{\cal C}(p) + (1\leftrightarrow2) \Big] ,\nonumber\\
&& J^{\mu\nu}_{\rm V} = 2T\sum_m\int\frac{d^3p}{(2\pi)^3} {\rm tr}\Big[ \gamma^\mu\gamma_5 S_1(p')\gamma^\nu\gamma_5 S_2^{\cal C}(p)\nonumber\\
&&\ \ \ \ \ \ \ \ \ \ \ \  + (1\leftrightarrow2) \Big] ,  \label{LoopFunctions}
\end{eqnarray}
within the imaginary-time formalism, where $p'=p+q$ and $p_0=i\omega_m$ [$\omega_m=(2m+1)\pi T$] and $q_0=i\bar{\omega}_n$ ($\bar{\omega}_n=2n\pi T$). The quark propagators are
\begin{eqnarray}
S_f(p) &=& {\rm F.T.}\langle0|{\rm T}\psi_f(x)\bar{\psi}_f(0)|0\rangle \nonumber\\
&=& i\sum_{\zeta={\rm p},{\rm a}}\frac{\Lambda^{(f)}_\zeta({\bm p})}{p_0-E^{(f)}_{\bm p}}\gamma_0 \ \ \ (f=1\ {\rm or}\ 2),
\end{eqnarray}
with the symbol ``F.T.'' standing for the Fourier transformation and $\Lambda_\zeta^{(f)}({\bm p})$ is a projection operator for the positive-energy and negative-energy states:
\begin{eqnarray}
\Lambda^{(f)}_{\zeta}({\bm p}) &=& \frac{E_{\bm p}^{(f)}+\eta_\zeta(M_f\gamma_0-{\bm p}\cdot{\bm \alpha})}{2E^{(f)}_{\bm p}} .
\end{eqnarray}
$E_{\bm p}^{(f)} = \sqrt{{\bm p}^2+M_f^2}$ is the single-particle energy of the quark $f$, and $\eta_{{\rm p}/{\rm a}}=\pm1$.
In addition, as explained in the main text the scalar diquarks can mix with the vector diquarks through a derivative coupling when their flavor structures are identical. Such a process is mediated by the following transition loop function:
\begin{eqnarray}
J^{\mu}_{{\rm S-V}}(q) &=& 2T\sum_m\int\frac{d^3p}{(2\pi)^3} {\rm tr}\Big[\gamma_5 S_1(p')\gamma^\mu\gamma_5 S_2^{\cal C}(p) \nonumber\\
&&\ \ \ \ \ \ \ \ \ \ + (1\leftrightarrow2)\Big] .
\end{eqnarray}
Then, calculating the kinetic parts
\begin{eqnarray}
T_{\rm S}^{\zeta'\zeta} &\equiv& {\rm tr}[\gamma_5\Lambda^{(1)}_{\zeta'}({\bm p}')\gamma_5\Lambda^{(2)}_\zeta({\bm p})] \nonumber\\
&=&1 + \frac{\eta_{\zeta'}\eta_\zeta({\bm p}'\cdot{\bm p}+M_1M_2)}{E_{{\bm p}'}^{(1)}E_{\bm p}^{(2)}} ,
\end{eqnarray}
\begin{eqnarray}
T_{\rm PS}^{\zeta'\zeta} &\equiv& {\rm tr}[\Lambda^{(1)}_{\zeta'}({\bm p}')\Lambda^{(2)}_\zeta({\bm p})] \nonumber\\
&=&1 + \frac{\eta_{\zeta'}\eta_\zeta({\bm p}'\cdot{\bm p}-M_1M_2)}{E_{{\bm p}'}^{(1)}E_{\bm p}^{(2)}} ,
\end{eqnarray}
\begin{widetext}
\begin{eqnarray}
T_{\rm AV}^{\zeta'\zeta,\mu\nu} &\equiv& {\rm tr}[\gamma^\mu\Lambda^{(1)}_{\zeta'}({\bm p}')\gamma_5\Lambda^{(2)}_\zeta({\bm p})] \nonumber\\
&=& \frac{1}{E_{{\bm p}'}^{(1)}E_{\bm p}^{(2)}} \Big[E_{{\bm p}'}^{(1)}E_{\bm p}^{(2)}(2g^{\mu0}g^{\nu0}-g^{\mu\nu}) + \eta_\zeta E^{(1)}_{{\bm p}'}p^j(g^{\mu0}g^{\nu j}+g^{\mu j}g^{\nu0}) - \eta_{\zeta'}E^{(2)}_{\bm p}p'^i (g^{\mu i}g^{\nu0} + g^{\mu0}g^{\nu i}) \nonumber\\
&&-\eta_{\zeta'}\eta_\zeta p'^ip^j(g^{\mu i}g^{\nu j}- g^{\mu\nu}g^{ij}+g^{\mu j}g^{i\nu})-\eta_{\zeta'}\eta_\zeta M_1M_2 g^{\mu\nu}\Big] ,
\end{eqnarray}
\begin{eqnarray}
T_{\rm V}^{\zeta'\zeta,\mu\nu} &\equiv& {\rm tr}[\gamma^\mu\Lambda^{(1)}_{\zeta'}({\bm p}')\gamma_5\Lambda^{(2)}_\zeta({\bm p})] \nonumber\\
&=& \frac{1}{E_{{\bm p}'}^{(1)}E_{\bm p}^{(2)}} \Big[E_{{\bm p}'}^{(1)}E_{\bm p}^{(2)}(2g^{\mu0}g^{\nu0}-g^{\mu\nu}) + \eta_\zeta E^{(1)}_{{\bm p}'}p^j(g^{\mu0}g^{\nu j}+g^{\mu j}g^{\nu0}) - \eta_{\zeta'}E^{(2)}_{\bm p}p'^i (g^{\mu i}g^{\nu0} + g^{\mu0}g^{\nu i}) \nonumber\\
&& -\eta_{\zeta'}\eta_\zeta p'^ip^j(g^{\mu i}g^{\nu j}- g^{\mu\nu}g^{ij}+g^{\mu j}g^{i\nu})+\eta_{\zeta'}\eta_\zeta M_1M_2 g^{\mu\nu}\Big] ,
\end{eqnarray}
\begin{eqnarray}
T_{\rm S-V}^{\zeta'\zeta,\mu} &\equiv& {\rm tr}[\gamma_5\Lambda^{(1)}_{\zeta'}({\bm p'})\gamma^\mu\gamma_5\Lambda_\zeta^{(2)}({\bm p})] \nonumber\\
&=& \frac{1}{E_{{\bm p}'}^{(1)}E_{\bm p}^{(2)}}\Bigg[(\eta_{\zeta'}E^{(2)}_{\bm p} M_1+ \eta_\zeta E^{(1)}_{{\bm p}'}M_2) g^{\mu0} - \eta_{\zeta'}\eta_\zeta(M_2 p'^i-M_1 p^i)g^{\mu i}\Bigg] ,
\end{eqnarray}
and employing the Matsubara summation formula
\begin{eqnarray}
T\sum_m\frac{1}{(i\omega_m+i\bar{\omega}_n-\epsilon')(i\omega_m-\epsilon)} = \frac{f_F(\epsilon)-f_F(\epsilon')}{i\bar{\omega}_n-\epsilon'+\epsilon} ,
\end{eqnarray}
with $f_F(\epsilon)=1/({\rm e}^{\epsilon/T}+1)$ being the Fermi-Dirac distribution function, the respective one-loop functions are evaluated in terms of the three-dimensional integral form as follows:
\begin{eqnarray}
 {\cal J}_{\rm S}  = -4\int\frac{d^3p}{(2\pi)^3}\sum_{\zeta',\zeta={\rm p},{\rm a}}\left(1+\frac{\eta_{\zeta'}\eta_\zeta({\bm p}_+\cdot{\bm p}_-+M_1M_2)}{E_{{\bm p}_+}^{(1)}E_{{\bm p}_-}^{(2)}}\right) \frac{1-f_F\left(\eta_{\zeta}E_{{\bm p}_-}^{(2)}\right)-f_F\left(\eta_{\zeta'}E_{{\bm p}_+}^{(1)}\right)}{q_0-\eta_{\zeta'}E_{{\bm p}_+}^{(1)}-\eta_\zeta E_{{\bm p}_-}^{(2)}}  , \label{JSApp}
\end{eqnarray}
\begin{eqnarray}
 {\cal J}_{\rm PS} =  -4\int\frac{d^3p}{(2\pi)^3}\sum_{\zeta',\zeta={\rm p},{\rm a}}\left(1+\frac{\eta_{\zeta'}\eta_\zeta({\bm p}_+\cdot{\bm p}_--M_1M_2)}{E_{{\bm p}_+}^{(1)}E_{{\bm p}_-}^{(2)}}\right) \frac{1-f_F\left(\eta_{\zeta}E_{{\bm p}_-}^{(2)}\right)-f_F\left(\eta_{\zeta'}E_{{\bm p}_+}^{(1)}\right)}{q_0-\eta_{\zeta'}E_{{\bm p}_+}^{(1)}-\eta_\zeta E_{{\bm p}_-}^{(2)}} , \label{JPSApp}
 \end{eqnarray}
 \begin{eqnarray}
&&{\cal J}_{\rm AV}^{00} =- 4\int\frac{d^3p}{(2\pi)^3}\sum_{\zeta',\zeta={\rm p}, {\rm a}}\left(1-\frac{\eta_{\zeta'}\eta_\zeta({\bm p}_+\cdot{\bm p}_-+M_1M_2)}{E_{{\bm p}_+}^{(1)}E_{{\bm p}_-}^{(2)}}\right)  \frac{1-f_F\left(\eta_{\zeta}E_{{\bm p}_-}^{(2)}\right)-f_F\left(\eta_{\zeta'}E_{{\bm p}_+}^{(1)}\right)}{q_0-\eta_{\zeta'}E_{{\bm p}_+}^{(1)}-\eta_\zeta E_{{\bm p}_-}^{(2)}}, \nonumber\\
&&{\cal J}_{\rm AV}^{0j} = -4\int\frac{d^3p}{(2\pi)^3}\sum_{\zeta',\zeta={\rm p}, {\rm a}}\left(\frac{\eta_{\zeta'}p_+^j}{E_{{\bm p}_+}^{(1)}}-\frac{\eta_{\zeta}p_-^j}{E_{{\bm p}_-}^{(2)}}\right) \frac{1-f_F\left(\eta_{\zeta}E_{{\bm p}_-}^{(2)}\right)-f_F\left(\eta_{\zeta'}E_{{\bm p}_+}^{(1)}\right)}{q_0-\eta_{\zeta'}E_{{\bm p}_+}^{(1)}-\eta_\zeta E_{{\bm p}_-}^{(2)}}, \nonumber\\
&&{\cal J}_{\rm AV}^{ij} =- 4\int\frac{d^3p}{(2\pi)^3}\sum_{\zeta',\zeta={\rm p}, {\rm a}}\Bigg(\delta^{ij}-\frac{\eta_{\zeta'}\eta_\zeta (p_+^ip_-^j+p_+^jp_-^i)}{E_{{\bm p}_+}^{(1)}E_{{\bm p}_-}^{(2)}}  +\frac{\eta_{\zeta'}\eta_\zeta({\bm p}_+\cdot{\bm p}_-+M_1M_2)\delta^{ij}}{E_{{\bm p}_+}^{(1)}E_{{\bm p}_-}^{(2)}}\Bigg)  \frac{1-f_F\left(\eta_{\zeta}E_{{\bm p}_-}^{(2)}\right)-f_F\left(\eta_{\zeta'}E_{{\bm p}_+}^{(1)}\right)}{q_0-\eta_{\zeta'}E_{{\bm p}_+}^{(1)}-\eta_\zeta E_{{\bm p}_-}^{(2)}} , \nonumber\\ \label{JijAV}
\end{eqnarray}
 \begin{eqnarray}
&&{\cal J}_{\rm V}^{00} = -4\int\frac{d^3p}{(2\pi)^3}\sum_{\zeta',\zeta={\rm p}, {\rm a}}\left(1-\frac{\eta_{\zeta'}\eta_\zeta({\bm p}_+\cdot{\bm p}_--M_1M_2)}{E_{{\bm p}_+}^{(1)}E_{{\bm p}_-}^{(2)}}\right)  \frac{1-f_F\left(\eta_{\zeta}E_{{\bm p}_-}^{(2)}\right)-f_F\left(\eta_{\zeta'}E_{{\bm p}_+}^{(1)}\right)}{q_0-\eta_{\zeta'}E_{{\bm p}_+}^{(1)}-\eta_\zeta E_{{\bm p}_-}^{(2)}}, \nonumber\\
&&{\cal J}_{\rm V}^{0j} =- 4\int\frac{d^3p}{(2\pi)^3}\sum_{\zeta',\zeta={\rm p}, {\rm a}}\left(\frac{\eta_{\zeta'}p_+^j}{E_{{\bm p}_+}^{(1)}}-\frac{\eta_{\zeta}p_-^j}{E_{{\bm p}_-}^{(2)}}\right) \frac{1-f_F\left(\eta_{\zeta}E_{{\bm p}_-}^{(2)}\right)-f_F\left(\eta_{\zeta'}E_{{\bm p}_+}^{(1)}\right)}{q_0-\eta_{\zeta'}E_{{\bm p}_+}^{(1)}-\eta_\zeta E_{{\bm p}_-}^{(2)}}, \nonumber\\
&&{\cal J}_{\rm V}^{ij} =- 4\int\frac{d^3p}{(2\pi)^3}\sum_{\zeta',\zeta={\rm p}, {\rm a}}\Bigg(\delta^{ij}-\frac{\eta_{\zeta'}\eta_\zeta (p_+^ip_-^j+p_+^jp_-^i)}{E_{{\bm p}_+}^{(1)}E_{{\bm p}_-}^{(2)}}  +\frac{\eta_{\zeta'}\eta_\zeta({\bm p}_+\cdot{\bm p}_--M_1M_2)\delta^{ij}}{E_{{\bm p}_+}^{(1)}E_{{\bm p}_-}^{(2)}}\Bigg)  \frac{1-f_F\left(\eta_{\zeta}E_{{\bm p}_-}^{(2)}\right)-f_F\left(\eta_{\zeta'}E_{{\bm p}_+}^{(1)}\right)}{q_0-\eta_{\zeta'}E_{{\bm p}_+}^{(1)}-\eta_\zeta E_{{\bm p}_-}^{(2)}} , \nonumber\\ \label{JVComp}
\end{eqnarray}
and
\begin{eqnarray}
&&{\cal J}_{\rm S-V}^{0} = 4\int\frac{d^3p}{(2\pi)^3}\sum_{\zeta',\zeta={\rm p}, {\rm a}}\left(\frac{\eta_{\zeta'}M_1}{E_{{\bm p}_+}^{(1)}}+\frac{\eta_{\zeta}M_2}{E_{{\bm p}_-}^{(2)}}\right) \frac{1-f_F\left(\eta_{\zeta}E_{{\bm p}_-}^{(2)}\right)-f_F\left(\eta_{\zeta'}E_{{\bm p}_+}^{(1)}\right)}{q_0-\eta_{\zeta'}E_{{\bm p}_+}^{(1)}-\eta_\zeta E_{{\bm p}_-}^{(2)}} , \nonumber\\
&&{\cal J}_{\rm S-V}^{i} = 4\int\frac{d^3p}{(2\pi)^3}\sum_{\zeta',\zeta={\rm p}, {\rm a}}\frac{\eta_{\zeta'}\eta_\zeta(M_2p_+^i-M_1p_-^i)}{E_{{\bm p}_+}^{(1)}E_{{\bm p}_-}^{(2)}} \frac{1-f_F\left(\eta_{\zeta}E_{{\bm p}_-}^{(2)}\right)-f_F\left(\eta_{\zeta'}E_{{\bm p}_+}^{(1)}\right)}{q_0-\eta_{\zeta'}E_{{\bm p}_+}^{(1)}-\eta_\zeta E_{{\bm p}_-}^{(2)}} , \label{JSVApp}
\end{eqnarray}
\end{widetext}
where the three-dimensional momentum has been shifted such that the integrands are symmetrically expressed by ${\bm p}_\pm={\bm p}\pm{\bm q}/2$.

The above one-loop functions enable us to evaluate the scattering amplitudes ${\cal T}=({\cal K}^{-1}-{\cal J})^{-1}$ for the respective channels and their masses. Among them, in particular, the mass of scalar diquark is evaluated by a pole position of the amplitude at rest with a $2\times2$ kernel matrix
\begin{eqnarray}
{\cal K} = \left(
\begin{array}{cc}
{\cal K}_{\rm S} & 0 \\
0 & {\cal K}_{\rm V} \\
\end{array}
\right) 
\end{eqnarray}
and a loop-function matrix
\begin{eqnarray}
{\cal J} = \left(
\begin{array}{cc}
{\cal J}_{\rm S} & {\cal J}_{\rm S-V}^0 \\
{\cal J}^0_{\rm S-V} & {\cal J}_{\rm V}^{00} \\
\end{array}
\right) ,
\end{eqnarray}
where ${\cal K}_{\rm S}$ and ${\cal K}_{\rm V}$ are the kernels for scalar and vector diquarks, respectively.

\section{Ward-Takahashi identity}
\label{sec:GaugeInvariance}

The one-loop functions of spin-$1$ diquarks are essentially the same as the (axial)vector current-current correlators, so that they should satisfy the WTIs derived from the appropriate current (non)conservation laws. Here we show their derivations. 

As in Appendix~\ref{sec:LoopFunctions}, let us start with a fermionic theory with two flavors $\psi=(\psi_1,\psi_2)$, the masses of which are different:
\begin{eqnarray}
{\cal L}_{\rm toy} = \bar{\psi}(i\Slash{\partial}-{\cal M})\psi + {\cal L}_{I} ,\label{Ltoy}
\end{eqnarray}
where ${\cal M} = {\rm diag}(M_1,M_2)$. The second term ${\cal L}_{I}$ is an arbitrary interaction Lagrangian which is invariant under a flavor $U(2)$ transformation: $\psi\to {\rm exp}(-i\theta^a \tau^a)\psi$ with $\tau^a$ being $U(2)$ generators: $\tau^0={\bm 1}_{2\times2}$ and $\tau^{a=1,2,3}$ is the Pauli matrix. When we consider an infinitesimal rotation generated by $\tau^1$ the Lagrangian~(\ref{Ltoy}) transforms as $\delta {\cal L}_{\rm toy} =-i\theta^1\bar{\psi}[\tau^1,{\cal M}]\psi =\theta^1(M_2-M_1)\bar{\psi}\tau^2\psi$. Thus, from the Noether's theorem the corresponding current (non)conservation law reads
\begin{eqnarray}
\partial^\mu j_{V,\mu}^1 = (M_2-M_1)\bar{\psi}\tau^2\psi, \ {\rm with}\ j_{V,\mu}^1 = \bar{\psi}\gamma_\mu\tau^1\psi .
\end{eqnarray}
In a similar way, under an axial transformation generated by $\psi \to {\rm exp}(-i\theta^1\gamma_5\tau^1)\psi$ the Noether's theorem yields
\begin{eqnarray}
\partial^\mu j_{A,\mu}^1 = i(M_1+M_2)\bar{\psi}\tau^1\gamma_5\psi, \ {\rm with}\ j_{A,\mu}^1 = \bar{\psi}\gamma_\mu\gamma_5\tau^1\psi . \nonumber\\
\end{eqnarray}
Those current (non)conservation laws hold at quantum level as well due to the absence of quantum anomalies.

From Eq.~(\ref{LoopFunctions}) the loop functions ${\cal J}^{\mu\nu}_{\rm V}$ and ${\cal J}^{\mu\nu}_{\rm A}$ are, at least within one-loop level, expressed in terms of the above currents as
\begin{eqnarray}
{\cal J}^{\mu\nu}_{\rm AV} &=& 2i {\rm F.T.}\langle0|{\rm T}j_{V}^{1,\mu}(x)j_{V}^{1,\nu}(0)|0\rangle , \nonumber\\
{\cal J}^{\mu\nu}_{\rm V} &=& 2i {\rm F.T.}\langle0|{\rm T}j_{A}^{1,\mu}(x)j_{A}^{1,\nu}(0)|0\rangle ,
\end{eqnarray}
respectively. Therefore, with the help of the equal-time anticommutation relation of $\psi$ and $\bar{\psi}$, one can obtain the following WTIs
\begin{eqnarray}
q_\mu{\cal J}^{\mu\nu}_{\rm AV} &=& \Delta_{\rm AV}^\nu , \nonumber\\
q_\mu{\cal J}^{\mu\nu}_{\rm V} &=& \Delta_{\rm V}^\nu , \label{WTIs}
\end{eqnarray}
where $\Delta_{\rm AV}^\nu$ and $\Delta_{\rm V}^\nu$ are evaluated to be
\begin{eqnarray}
\Delta_{\rm AV}^\nu &=& 2i {\rm F.T.}\langle0|{\rm T}i\partial^x_\mu j_V^{1,\mu}(x)j_V^{1,\nu}(0)|0\rangle \nonumber\\
&=&2(M_1-M_2) {\rm F.T.}\langle0|T\bar{\psi}(x)\tau^2\psi(x)j_V^{1,\nu}(0)|0\rangle , \nonumber\\
&=& 2(M_1-M_2)T\sum_m\int\frac{d^3p}{(2\pi)^3} \nonumber\\
&\times&{\rm tr}\Big[ S^{(1)}(p')\gamma^\nu S^{(2)}(p)  - (1\leftrightarrow 2)\Big] , \label{DeltaAV}
\end{eqnarray}
and 
\begin{eqnarray}
\Delta_{\rm V}^\nu &=& 2i {\rm F.T.}\langle0|{\rm T}i\partial^x_\mu j_A^{1,\mu}(x)j_A^{1,\nu}(0)|0\rangle \nonumber\\
&=&-2i(M_1+M_2) {\rm F.T.}\langle0|T\bar{\psi}(x)\tau^1\psi(x)j_A^{1,\nu}(0)|0\rangle , \nonumber\\
&=& -2(M_1+M_2) T\sum_m\int\frac{d^3p}{(2\pi)^3} \nonumber\\
&\times&{\rm tr}\Big[ \gamma_5 S^{(1)}(p')\gamma^\nu\gamma_5 S^{(2)}(p)  + (1\leftrightarrow 2)\Big] ,  \label{DeltaV}
\end{eqnarray}
respectively. In the last equalities we have adopted one-loop approximations. Equation~(\ref{DeltaAV}) implies that when $M_1=M_2$, $\Delta_{\rm AV}^\nu$ is reduced to zero and $q_\mu {\cal J}_{\rm AV}^{\mu\nu}=0$ follows. This is because the Lagrangian~(\ref{Ltoy}) possesses the exact flavor $U(2)$ symmetry in this case and the corresponding current is conserved. On the other hand, even when $M_1=M_2$, $\Delta_V^\nu$ does not converge to zero since the mass term in Eq.~(\ref{Ltoy}) always breaks the axial $U(2)$ symmetry.

Within the three-dimensional-integral form the current-nonconservation measures $\Delta_{\rm AV}^\nu$ and $\Delta_{\rm V}^\nu$ are calculated to be 
\begin{widetext}
\begin{eqnarray}
\Delta_{\rm AV}^\nu &=& 4(M_2-M_1)\int\frac{d^3p}{(2\pi)^3}\sum_{\zeta',\zeta={\rm p,a}}\left[\left(\frac{\eta_{\zeta'}M_1}{E_{{\bm p}_+}^{(1)}}-\frac{\eta_{\zeta}M_2}{E^{(2)}_{{\bm p}_-}}\right)g^{\nu0}+\frac{\eta_{\zeta'}\eta_\zeta(M_1 p_-^i+M_2p_+^i)g^{\nu i}}{E_{{\bm p}_+}^{(1)}E_{{\bm p}_-}^{(2)}}\right] \nonumber\\
&& \times \frac{1-f_F\left(\eta_{\zeta}E_{{\bm p}_-}^{(2)}\right)-f_F\left(\eta_{\zeta'}E_{{\bm p}_+}^{(1)}\right)}{q_0-\eta_{\zeta'}E_{{\bm p}_+}^{(1)}-\eta_\zeta E_{{\bm p}_-}^{(2)}} , \nonumber\\
\Delta_{\rm V}^\nu &=& -4(M_1+M_2)\int\frac{d^3p}{(2\pi)^3}\sum_{\zeta',\zeta={\rm p,a}}\left[\left(\frac{\eta_{\zeta'}M_1}{E_{{\bm p}_+}^{(1)}}+\frac{\eta_{\zeta}M_2}{E^{(2)}_{{\bm p}_-}}\right)g^{\nu0}+\frac{\eta_{\zeta'}\eta_\zeta(M_1 p_-^i-M_2p_+^i)g^{\nu i}}{E_{{\bm p}_+}^{(1)}E_{{\bm p}_-}^{(2)}}\right] \nonumber\\
&& \times \frac{1-f_F\left(\eta_{\zeta}E_{{\bm p}_-}^{(2)}\right)-f_F\left(\eta_{\zeta'}E_{{\bm p}_+}^{(1)}\right)}{q_0-\eta_{\zeta'}E_{{\bm p}_+}^{(1)}-\eta_\zeta E_{{\bm p}_-}^{(2)}} , \label{DeltaThree}
\end{eqnarray}
\end{widetext}
respectively. Then, from the loop functions derived in Appendix~\ref{sec:LoopFunctions} and Eq.~(\ref{DeltaThree}), for $\nu=0$ one can numerically show that the WTIs in Eq.~(\ref{WTIs}) are indeed satisfied by introducing a three-dimensional cutoff. On the other hand, when $\nu=j$ the identities are found not to be held, implying that within the three-dimensional-cutoff form the regularization artifacts contaminate the spatial parts. As another consequence of this partial violation, one can see that the quadratic divergences only emerge in ${\cal J}_{\rm AV}^{ij}$ and ${\cal J}_{\rm V}^{ij}$. It should be noted that the temperature-dependent parts satisfy the WTIs since their contributions are always finite due to the Boltzmann suppression of high-momentum modes. Indeed, we have checked it with naive three-dimensional cutoffs by numerically confirming the identity 
\begin{eqnarray}
q^jq_\mu{\cal J}_{\rm AV}^{\mu j}\Big|_{T.\, {\rm dep.}} &=& q^j\Delta_{\rm AV}^j\Big|_{T.\, {\rm dep.}} , \nonumber\\
q^jq_\mu{\cal J}_{\rm V}^{\mu j}\Big|_{T.\, {\rm dep.}} &=& q^j\Delta_{\rm V}^j\Big|_{T.\, {\rm dep.}} ,  \label{TDepWTI}
\end{eqnarray}
 derived from Eq.~(\ref{WTIs}).

It would be pedagogical to check the WTIs~(\ref{WTIs}) analytically by employing a certain regularization, then at the end of this appendix we calculate the loop functions ${\cal J}^{\mu\nu}_{\rm AV(V)}$ and current-nonconservation measures $\Delta_{\rm AV(V)}^\nu$ with the dimensional regularization. Here, we restrict ourselves in the vacuum since only the diverging vacuum parts are nontrivial. Within the dimensional regularization, ${\cal J}^{\mu\nu}_{\rm AV}$, ${\cal J}^{\mu\nu}_{\rm V}$, $\Delta_{\rm AV}^\nu$ and $\Delta_{\rm V}^\nu$ are evaluated to be
\begin{eqnarray}
{\cal J}^{\mu\nu}_{\rm AV} &=& 32\int_0^1 dxX_{12}(x^2-x)(q^2g^{\mu\nu}-q^\mu q^\nu) \nonumber\\
&+& 16\int_0^1 dx X_{12} (M_1-M_2)\Big[(1-x)M_1-xM_2\Big]g^{\mu\nu}, \nonumber\\
{\cal J}^{\mu\nu}_{\rm V} &=& 32\int_0^1 dxX_{12}(x^2-x)(q^2g^{\mu\nu}-q^\mu q^\nu) \nonumber\\
&+& 16\int_0^1 dx X_{12}(M_1+M_2)\Big[(1-x)M_1+xM_2\Big] g^{\mu\nu}, \nonumber\\
\Delta_{\rm AV}^\nu &=& 16\int_0^1 dx X_{12} (M_1-M_2)\Big[(1-x)M_1-xM_2\Big]q^{\nu}, \nonumber\\
\Delta_{\rm V}^\nu &=& 16\int_0^1 dx X_{12} (M_1+M_2)\Big[(1-x)M_1+xM_2\Big]q^{\nu}, \nonumber\\
\end{eqnarray}
where the logarithmic divergences are expressed as the following $d\to4$ pole:
\begin{eqnarray}
X_{12} \equiv \frac{1}{(4\pi)^{d/2}}\frac{\Gamma\left[2-\frac{d}{2}\right]}{\Delta_{12}^{2-d/2}} , \label{XxDef}
\end{eqnarray}
with $\Delta_{12} = (1-x)M_1^2+xM_2^2-x(1-x)q^2$. Note that the quadratic divergences signaled by $d\to2$ poles have been eliminated. Therefore, one can easily verify that the WTIs~(\ref{WTIs}) indeed hold for both the axialvector- and vector-diquark channels.

\section{Gauge invariant expressions with three-dimensional cutoffs}
\label{sec:GaugeInvariantJs}

In the present paper, we have employed the loop functions in Eqs.~(\ref{JSApp}) -~(\ref{JSVApp}) which contain quadratic divergences with the three-dimensional proper-time regularization to eliminate the imaginary parts. As explained in the main text, this procedure is suitable for treating the confined diquarks inside SHBs although it also leads to violation of the gauge invariance. However, it would be helpful for readers to present the loop functions which is capable of respecting the WTIs with naive three-dimensional cutoffs, thus, here we derive them. We note that in this appendix we stick to zero temperature.

Since the three-dimensional-integral form breaks the Lorentz invariance even at zero temperature, the loop function generally takes the form of
\begin{eqnarray}
{\cal J}^{\mu\nu}_v &=& A_v(g^{\mu\nu}-v^\mu v^\nu)+B_v(q^2g^{\mu\nu}-q^\mu q^\nu) \nonumber\\
&& +C_vg^{\mu\nu} + D_v(v^\mu q^\nu+v^\nu q^\mu) , \label{JMuNuGeneral}
\end{eqnarray}
with $v=AV$ or $V$. In this decomposition $v^{\mu} = g^{\mu0}= (1,{\bm 0})$ is introduced to separate the time and spatial components explicitly. As explained in Appendix~\ref{sec:GaugeInvariance} the time components of ${\cal J}_v^{\mu\nu}$ always respect the WTIs: $q_\mu{\cal J}_v^{\mu0}=\Delta_v^0$, within the present three-dimensional cutoff scheme. Hence, by multiplying $q_\mu$ from the both sides of Eq.~(\ref{JMuNuGeneral}) and setting $\nu=0$, one can find the following constraint for $C_v$:
\begin{eqnarray}
C_v = \frac{1}{q_0}\Delta^0_v-\frac{q_0^2+q^2}{q_0}D_v .
\end{eqnarray}
Inserting this into Eq.~(\ref{JMuNuGeneral}) yields
\begin{eqnarray}
{\cal J}_v^{\mu\nu} &=&  A_v(g^{\mu\nu}-v^\mu v^\nu)+B_v(q^2g^{\mu\nu}-q^\mu q^\nu) \nonumber\\
&+& D_v\left[( v^\mu q^\nu+v^\nu q^\mu)-\frac{q_0^2+q^2}{q_0}g^{\mu\nu}\right] +\frac{ \Delta_v^{0}}{q_0} g^{\mu\nu} . \nonumber\\
\end{eqnarray}
Then, the $\nu=0$ component reads
\begin{eqnarray}
{\cal J}_v^{\mu0} &=& \left(B_v-\frac{D_v}{q_0}\right)(q^2g^{\mu0}-q^\mu q^0) + \frac{ \Delta_{v}^{ 0}}{q_0} g^{\mu0} . \label{JvMu0}
\end{eqnarray}
When we further take $\mu=0$ in this equation, 
\begin{eqnarray}
{\cal J}^{00}_v = -\left(B_v-\frac{D_v}{q_0}\right){\bm q}^2 + \frac{ \Delta_{v}^{0}}{q_0}
\end{eqnarray}
is found, so Eq.~(\ref{JvMu0}) can be written in terms of only ${\cal J}_v^{00}$ and $\Delta_v^0$ as
\begin{eqnarray}
{\cal J}_v^{\mu 0} &=& -\frac{1}{{\bm q}^2}\left({\cal J}_v^{00}-\frac{\Delta^0_v}{q_0}\right)(q^2g^{\mu0}-q^\mu q^0)+ \frac{\Delta_v^0}{q_0}g^{\mu0}. \nonumber\\ \label{JvMu0_2}
\end{eqnarray}

Here, since ${\cal J}_v^{\mu0}$ does not violate the gauge invariance, one can infer that in the vacuum Lorentz covariant extension of Eq.~(\ref{JvMu0_2}) is a ``correct'' expression of the loop function. Therefore, by recovering the Lorentz covariance,
\begin{eqnarray}
\bar{\cal J}_v^{\mu \nu} &\equiv& -\frac{1}{{\bm q}^2}\left({\cal J}_v^{00}-\frac{\Delta^0_v}{q_0}\right)(q^2g^{\mu\nu}-q^\mu q^\nu)+ \frac{\Delta_v^0}{q_0}g^{\mu\nu}, \nonumber\\ \label{JvMuNuCorrect1}
\end{eqnarray}
or from $q_\mu{\cal J}_v^{\mu0}=\Delta_v^0$
\begin{eqnarray}
\bar{\cal J}_v^{\mu \nu} &=& \frac{1}{{\bm q}^2}\frac{q_i{\cal J}_v^{i0}}{q_0}(q^2g^{\mu\nu}-q^\mu q^\nu)+ \frac{\Delta_v^0}{q_0}g^{\mu\nu}, \nonumber\\ \label{JvMuNuCorrect2}
\end{eqnarray}
can be a plausible gauge-invariant loop function within the three-dimensional cutoff scheme. Equation~(\ref{JvMuNuCorrect1}) or (\ref{JvMuNuCorrect2}) implies that the correct loop function $\bar{\cal J}_v^{\mu\nu}$ can be evaluated by ${\cal J}_v^{00}$ (or ${\cal J}_v^{i0}$) and $\Delta_v^0$ solely and there is no need to compute the problematic ${\cal J}_v^{ij}$ as it should be.

\bibliography{reference}

\end{document}